\RequirePackage{fix-cm}
\documentclass[smallextended]{svjour3}       
\smartqed  
\usepackage{graphicx}
\usepackage{amssymb} 

\usepackage{subfigure}
\usepackage{amsmath}
\usepackage{amsfonts}
\usepackage{amsbsy}
\usepackage{times}
\usepackage{mathptmx}
\usepackage[left]{lineno}

 \journalname{J. Braz. Soc. Mech. Sci. Eng.}
\begin{document}

\title{Turbulent channel flow perturbed by triangular ripples\thanks{This is a pre-print of an article published in Journal of the Brazilian Society of Mechanical Sciences and Engineering, 40:138, 2018. The final authenticated version is available online at: http://dx.doi.org/10.1007/s40430-018-1050-7}
}


\author{Fernando David C\'u\~nez Benalc\'azar \and
				Gabriel Victor Gomes de Oliveira \and
        Erick de Moraes Franklin 
}


\institute{Fernando David C\'u\~nez Benalc\'azar \at
              School of Mechanical Engineering, University of Campinas - UNICAMP \\
							 \email{fernandodcb@fem.unicamp.br}               
									 \and
           Gabriel Victor Gomes de Oliveira \at
              School of Mechanical Engineering, University of Campinas - UNICAMP \\
							 \email{gabriel.vgo@gmail.com} 
										\and
					 Erick de Moraes Franklin (corresponding author)\at
              School of Mechanical Engineering, University of Campinas - UNICAMP \\
              Tel.: +55-19-35213375\\							
              orcid.org/0000-0003-2754-596X\\
              \email{franklin@fem.unicamp.br}             
}

\date{Received: date / Accepted: date}

\maketitle

\begin{abstract}
This paper presents an experimental investigation of the perturbation of a turbulent closed-conduit flow by two-dimensional triangular ripples. Two ripple configurations were employed: one single asymmetric triangular ripple, and two consecutive asymmetric triangular ripples, all of them with the same geometry. Different water flows were imposed over either one or two ripples fixed to the bottom wall of a closed conduit, and the flow field was measured by PIV (particle image velocimetry). Reynolds numbers based on the channel height were moderate, varying between 27500 and 34700. The regime was hydraulically smooth, and the blockage ratio was significant. Experimental data for this specific case remain scarce, and the physics involved has yet to be fully elucidated. Using the instantaneous flow fields, the mean velocities and fluctuations were computed, and the shear stresses over the ripples were determined. The velocities and stresses obtained in this way for the single ripple and for the pair of ripples are compared, and the surface shear stress is discussed in terms of bed stability. Our results show that for the single and upstream ripples the shear stress increases at the leading edge and decreases toward the crest, while for the downstream ripple it decreases monotonically between the reattachment point and the crest. The stress distribution over the downstream ripple, which is different from both the upstream and single ripples, is shown to sustain existing ripples over loose beds.
\end{abstract}

\section*{Nomenclature}
\label{section:nomenclature}

\subsection*{Roman symbols}
\begin{tabular}{l l}
$A$ & constant\\
$B$ & constant\\
$d$ & diameter (m)\\
$f$ & Darcy friction factor\\
$h$ & local height of the ripple (m)\\
$H$ & channel height (m)\\
$H_{eff}$ & distance between the PVC plates and the top wall of the channel (m)\\
$i$ & imaginary number\\
$k$ & wave number (m$^{-1}$)\\
$L$ & length scale (m)\\
$L_r$ & ripple total length (m)\\
$l_v$ & viscous length (m)\\
$P$ & turbulence production term (m$^2$/s$^3$)\\
$Q$ & volumetric flow rate (m$^3$/h)\\
$Re$ & Reynolds number based on the channel height\\
$Re_{dh}$ & Reynolds number based on the hydraulic diameter\\
$t$ & time (s)\\
$u$ & longitudinal component of the velocity (m/s)\\
$\overline{U}$ & cross-sectional mean velocity (m/s)\\
$u_*$ & shear velocity (m/s)\\
$v$ & vertical component of the velocity (m/s)\\
$\vec{V}$ & velocity vector (m/s)\\
$x$ & horizontal coordinate (m)\\
$y$ & vertical coordinate (m)\\
\end{tabular}

\subsection*{Greek symbols}
\begin{tabular}{l l}
$\Delta$ & perturbation field\\
$\kappa$ & von K\'arm\'an constant\\
$\lambda$ & wavelength (m)\\
$\mu$ & dynamic viscosity (Pa.s)\\
$\nu$ & kinematic viscosity (m$^2$/s)\\
$\rho$ & density (kg/m$^3$)\\
$\tau$ & shear stress (N/m$^2$)\\
$\theta$ & angle between the horizontal and the ripple surface ($^o$)\\
\end{tabular}

\subsection*{Subscripts}
\begin{tabular}{l l}
$0$ & relative to the flat surface (unperturbed flow)\\
$bla$ & relative to Blasius correlation\\
$d$ & relative to the displaced coordinate system\\
$dh$ & relative to the hydraulic diameter\\
$k$ & relative to Fourier space\\
$max$ & maximum value\\
$v$ & relative to the viscous layer\\
$\theta$ & aligned with the ripple surface\\
\end{tabular}

\subsection*{Superscripts}
\begin{tabular}{l l}
$+$  & normalized by the viscous length $l_v$ or by the shear velocity $u_{*,0}$\\
$\overline{\quad}$ & averaged in time\\
$'$ & fluctuation\\ 		
$\hat{\quad}$ & perturbation\\
\end{tabular}

\section{Introduction}

Turbulent flows over walls in the presence low hills are frequently found in industry and nature. Examples include air flows over desert dunes and ocean waves, water flows over sand ripples in rivers, and oil flows over sand ripples in petroleum pipelines. The perturbation of a boundary layer by triangular ripples introduces new scales in the problem, changing the velocity and stress distributions along the flow, quantities which are of importance for many engineering applications in both open and channel flows. Given its importance, many studies have been devoted to the perturbation of a two-dimensional turbulent boundary layer by a hill with a low aspect ratio \cite{Jackson_Hunt,Hunt_1,Belcher_Hunt,poggi2007}, or low hill (height-to-length ratio of ord(0.01), where ord stands for ``order of magnitude''). Of these studies, many use asymptotic methods and have helped to enhance understanding of the subject. When asymptotic methods are used, the turbulent boundary layer perturbed by a low hill is divided into a basic state and a perturbation, where the perturbation is assumed to be small compared with the basic flow. The fluid flow is therefore written as a basic unperturbed flow and a flow perturbation. For the shear stress on the bed surface

\begin{equation}
\tau\,=\,\tau_{0}(1\,+\,\hat{\tau})
\label{stress_total}
\end{equation}

\noindent where $\tau_{0}$ and $\hat{\tau}$ are the shear stresses on the bed caused by the basic flow and flow perturbation, respectively, the latter being dimensionless and $\ll 1$. In the case of a turbulent boundary layer on a flat wall, $\tau_{0} = \rho u_{*,0}^2$, where $ u_{*,0}$ is the shear velocity on the flat wall and $\rho$ is the density of the fluid. In addition, in asymptotic methods the turbulent boundary layer perturbed by a low hill is divided into two or more regions that can be used to determine the perturbed flow \cite{Belcher_Hunt}. In most cases, the lower regions are supposed to be in local equilibrium and turbulence models are employed.

The asymptotic methods used in \cite{Jackson_Hunt} and \cite{Hunt_1} are formally valid for hills with a low aspect ratio (ord(0.01)). However, Carruthers and Hunt \cite{Carruthers_Hunt} showed that good results can also be obtained when these methods are applied to shapes with aspect ratios of up to ord(0.1). In these cases, the perturbation must be computed for an envelope formed by the hill and the recirculation bubble that appears just downstream of the crest. Based on \cite{Jackson_Hunt,Hunt_1,Carruthers_Hunt}, Weng et al. \cite{Weng} computed the velocity perturbations up to the second order and applied the results to higher-aspect-ratio bedforms. The resulting expressions provide smoother matching between the regions.

Analyzing the results reported in \cite{Weng}, Kroy et al. \cite{Kroy_A,Kroy_C} grouped the terms responsible for the upstream shift of the perturbation and showed that only some dominant terms need to be considered. They proposed a simplified expression for the surface stress, maintaining the order of magnitude of both the amplitude and phase of the perturbation. They proposed that the dimensionless perturbation of the longitudinal shear stress in Fourier space is given by Eq. \ref{stress_pert_fourier}

\begin{equation}
\hat{\tau}_{k}=Ah(|k|+iBk)
\label{stress_pert_fourier}
\end{equation}

\noindent where $h$ is the local height of the hill, $k=2\pi\lambda^{-1}$ is the longitudinal wavenumber, $\lambda$ is the wavelength, $i$ is the imaginary number, and $A$ and $B$ are considered as constants. The first term in the parentheses is symmetric, as in the potential solution, and of $O$(1), where $O$ stands for ``order of the term'' in the asymptotic expansion. It is in phase with the bedform and is the main term responsible for the amplitude of the perturbation. The second term in the parentheses is antisymmetric and of $O$(2). Because it is out-of-phase with the bedform, it causes a small shift in the perturbation and changes its amplitude slightly. 

Andreotti et al. \cite{Andreotti_1,Andreotti_2} reviewed the morphology and dynamics of barchan (crescentic shape) dunes and proposed a simplified two-dimensional model. They presented the results of previous studies showing that the only unstable mechanism leading to the formation of dunes is the fluid flow, which must have an upstream shift with respect to the bedform. In addition, they proposed, based on their work and the results of \cite{Howard,Wiggs,Sauermann_2}, that the local shear velocity decreases at the beginning of the dune (leading edge) and then increases toward the crest, reaching at the crest a value approximately 1.4 times the value of the shear velocity over a flat bed.

Sauermann et al. \cite{Sauermann_3} measured the wind velocity and sand flux along a barchan dune in Jericoacoara, Brazil. The measurements were made at the vertical symmetry plane (central slice) of the dune, in the region upstream of its crest (windward side of the dune). The fluid flow was measured using a reference anemometer placed far upstream of the dune and another which was displaced along the dune. Both anemometers were placed 1 m above the surface. From the measured velocities, the authors computed the shear velocity by assuming a logarithmic profile. Their results show that the local shear velocity decreases slightly at the leading edge of the dune and then increases toward the crest to a value 1.4 times that over a flat bed. 

Poggi et al. \cite{poggi2007} performed experiments on turbulent water flows over a two-dimensional hilly surface at high Reynolds numbers. The experiments were performed in an 18 m-long, 0.90 m-wide, 1 m-deep flume with a sinusoidal bottom. The sinusoidal bedforms were 0.08 m high and 3.2 m long, and the Reynolds number used, which was based on the channel height, was $1.5 \cdot 10^5$. The longitudinal and vertical components of the flow field were measured by LDA (laser Doppler anemometry). The authors found that the analyses proposed in \cite{Jackson_Hunt} and \cite{Hunt_1} for isolated hills are also valid for a series of low hills with smooth shapes.

Using DNS (direct numerical simulation), Marquillie et al. \cite{Laval_1,Laval_2} investigated numerically the perturbation of a channel flow by a low hill with a smooth shape. The bedform used was smoother than a ripple or a dune, with no sharp angles and a downstream face with an angle well below the avalanche angle. However, with the Reynolds numbers used, a recirculation bubble was formed. The authors were interested in this configuration because it provides an adverse pressure gradient with a curved bottom wall and flat top wall. They found that the unfavorable pressure gradients downstream of the dune crest generate intense vortices near the crest region on both the bottom and top walls, increasing turbulence production at both walls. 

Franklin and Charru \cite{Franklin_9} and Charru and Franklin \cite{Franklin_10} studied the dynamics of barchan dunes and the perturbation they cause in closed-conduit water flows. The experiments were performed in a 6 m-long closed-conduit channel with a rectangular cross section (60 mm high and 120 mm wide) using Reynolds numbers between 9000 and 24000 based on channel height. The authors measured the fluid flow at the symmetry plane of moving and fixed barchan dunes by PIV (particle image velocimetry) and used the results to calculate the evolution of the shear stress along the dunes. They failed to find an upstream shift of the maximum shear stress with respect to the dune crest, a result that has yet to be explained.

Franklin and Ayek \cite{Franklin13} studied the perturbation of a fully developed turbulent boundary layer by a two-dimensional triangular hill. In their experiments, water flows were imposed over a triangular ripple with sharp edges fixed to the bottom wall of a closed-conduit channel. Confinement effects were present as the ripple was 8 mm high and the channel 43 mm high, corresponding to a blockage ratio of approximately $20\%$. The flow field was measured by PIV, and mean and fluctuation fields were computed. The authors found that at the leading edge of the ripple the $xy$ component of turbulent stresses increases to around eight times the value over the plane bed, while the viscous stress remains of the same order of magnitude, indicating that the inner regions of the perturbed flow are not in local equilibrium for the triangular ripple with sharp edges. Franklin and Ayek \cite{Franklin13} proposed that the absence of local equilibrium is due to the relatively large ratio between the vertical and longitudinal scales of the flow. However, it could be due to the sharp leading edge, which probably shed vortices for Reynolds numbers above some threshold. In addition, the authors found that the maximum shear stress on the triangular ripple occurs upstream of its crest, at a distance of approximately $50\%$ of the ripple length measured from the crest.

Few studies were devoted to the flow over two or more sequential ripples of triangular shape. Walker and Nickling \cite{Walker} performed experiments on turbulent air flows over either a single or a sequence of triangular ripples where they measured the surface stress using pressure transducers. The experiments were performed in a 0.92 m wide, 0.76 m high and 8 m long wind tunnel, and the employed ripples were 680 mm long and 80 mm high. The flow rates varied between 8 and 18 m/s, from which we can estimate $l_v/L_r$ = ord(10$^{-4}$), where $l_v$ is the viscous length and $L_r$ is the ripple length. The ripple length (ord(1 m)) and the air velocities (ord(10 m/s)) are much larger than the usually found for subaqueous ripples (ord(100 mm) and ord(1 m/s), respectively, \cite{Franklin_15}), and, consequently, the values of $l_v/L_r$ are much lower than the corresponding values for subaqueous ripples (ord(10$^{-3}$)). The authors did not measure the flow field, but measured the surface pressure on the ripples. From this, they computed the mean values and the fluctuations of the surface shear stress. However, the number of pressure ports and the acquisition frequency were small (only 19 ports for the sequence of ripples, acquiring at 100 Hz). For the isolated ripple, the authors found that the flow is stable over most of the ripple upwind face, with the exception of the leading edge and the crest regions, where stress fluctuations are higher and lower than expected, respectively. For the ripples within a sequence, the flow reattaches around 4.4$h$ to 8$h$ (352 to 640 mm) downstream of the crest of the upstream ripple, and downstream of this point the flow develops and becomes stable. They found that the stress caused by fluctuations increases at the leading edge and decays toward the crest for the single ripple, while it decays from the reattachment point toward the crest for the ripples within a sequence. For the mean shear stress, they found that it always increases toward the crest. We note that these behaviors may be different for subaqueous ripples because of their relative higher $l_v/L_r$ values.

Coleman et al. \cite{Coleman_3} performed experiments on turbulent water flows over two-dimensional dune profiles. The experiments were performed in a 0.44 m wide, 0.38 m high, 12 m long open channel, and the mean velocities varied between 5.0 and 5.9 m/s, corresponding to shear velocities between 0.0387 and 0.0424 m/s, and to Reynolds numbers of ord($10^5$). The channel bed consisted of a sequence of 15 two-dimensional dunes, each one with a shape similar to the symmetry plane of a barchan dune \cite{Franklin_9} and 0.75 m long, so that $l_v/L_r$ = ord(10$^{-5}$). The flow field was measured by acoustic Doppler velocimetry and the results are presented only in terms of mean velocities, from which the authors compute the shear stress at the bed by assuming logarithmic velocity profiles on the upstream faces of dunes. They found that the shear stress at the surface increases monotonically toward  the crest. If the logarithmic profiles can exist for subaqueous dunes with $l_v/L_r$ = ord(10$^{-5}$), this may be not true for smaller bedforms such as triangular ripples, for which $l_v/L_r$ = ord(10$^{-3}$).

This paper presents an experimental investigation on the perturbation of a closed-conduit flow by triangular ripples. The flow is turbulent, hydraulically smooth and under moderate Reynolds numbers (ord($10^4$)) with a high blockage ratio (around $20\%$). Two ripple configurations were employed: one single asymmetric triangular ripple, and two consecutive asymmetric triangular ripples, all of them having the same geometry. Unlike in \cite{Franklin13}, the ripples have slightly rounded edges to represent better the ripples found in nature. In the arrangement used here, water flows are imposed in a closed conduit with a rectangular cross section and either one or two ripples fixed to the bottom wall. There is a dearth of studies and, consequently, experimental data for this setup, and the physics associated with it has yet to be fully understood. One important question addressed in this paper is the distribution of the shear stress on the ripple surface for both isolated and a pair of ripples, which is necessary to understand the stability of these bedforms. The instantaneous flow fields are measured with PIV, and the results are used to determine the mean velocities and fluctuations. Shear stresses on the bed and turbulence production are investigated, and the surface shear stress is discussed in terms of bed stability. The stress distribution over the downstream ripple, which is different from both the upstream and single ripples, is shown to sustain existing ripples over loose beds.

\section{Experimental setup}
 
The experimental setup consisted of a water reservoir, two centrifugal pumps connected in parallel, an electromagnetic flow meter, a flow straightener, a 5 m-long channel with a rectangular cross section, a settling tank and a return line. The channel was a closed conduit of transparent material (plexiglass) with a 160 mm-wide, 50 mm-high cross section. The channel test section was 1 m long and started 3 m (40 hydraulic diameters) downstream of the channel inlet. The remaining 1 m section connected the test-section exit to the settling tank. The water flowed in a closed loop in the order described above.

The volutes and rotors of the centrifugal pumps were made of brass to avoid corrosion particles entering the water, as these can disturb PIV measurements. The flow rates were adjusted with a set of valves. The flow straightener, whose function was to homogenize the water flow, consisted of a divergent-convergent nozzle filled with glass spheres with a mean diameter, $d$, of 3 mm. 

\begin{figure}[ht]	
 	\centering
 	\includegraphics[width=0.9\columnwidth]{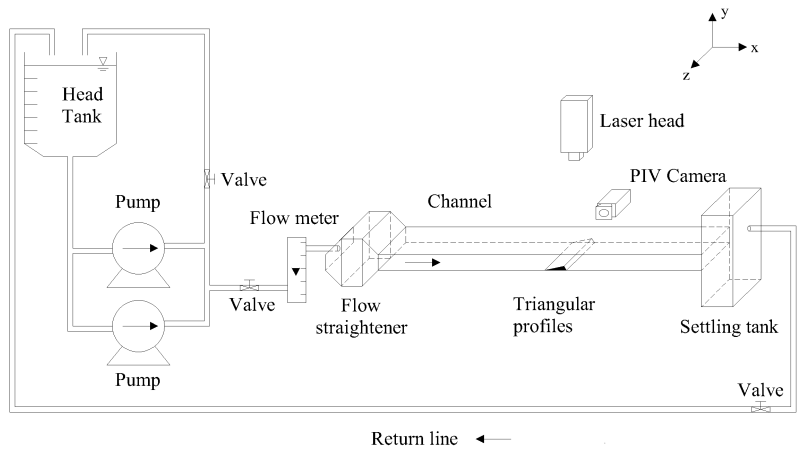}
 	\caption{Layout of the experimental setup.}
 	\label{fig:1}
\end{figure}

PVC plates 7 mm thick were placed in the channel so that they covered the entire bottom wall, reducing the channel height to 43 mm. A 99.4 mm-long, 9.4 mm-high triangular-shaped ripple with a 6.36$^{\circ}$ upstream angle was used to model each two-dimensional ripple. The edges of the ripple were made slightly rounded with a radius of curvature of approximately 0.2 mm, the same order of magnitude as the mean diameter of sand. The model ripple was of the same scale (including the edges) as the aquatic ripples found on, for example, river beds and the closed-conduit ripples encountered in industry.

\begin{figure}[ht]	
 	\centering
 	\includegraphics[width=0.75\columnwidth]{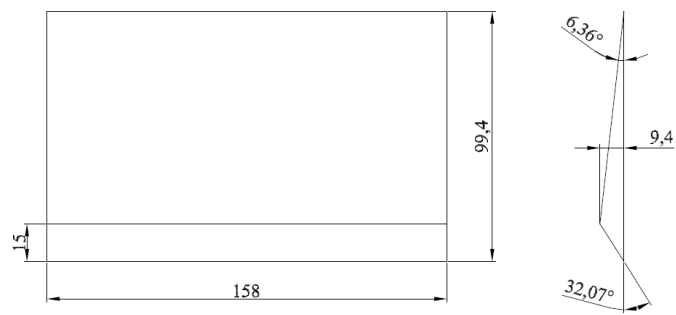}
 	\caption{Dimensions of each ripple.}
 	\label{fig:ripple}
\end{figure}

\begin{figure}[ht]	
	\centering
	\includegraphics[width=0.5\columnwidth]{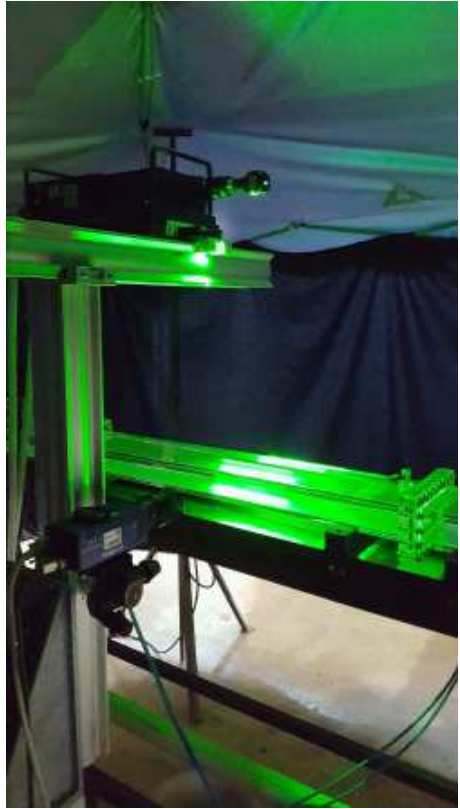}
	\caption{Photograph of the test section.}
	\label{fig:photo_test_section}
 \end{figure}

Three ripples were built for the two ripple configurations investigated in this paper: either one single asymmetric triangular ripple or two consecutive asymmetric triangular ripples were fixed to the PVC plate on the bottom wall of the test section. The ripples were made of black plastic (polyoxymethylene), and the PVC plates were painted in black to minimize undesirable reflections. Fig. \ref{fig:ripple} shows the dimensions of the model ripple. Fig. \ref{fig:photo_test_section} shows a photograph of the test section.

Two flow rates were used (8 and 10 m$^{3}$/h), corresponding to mean cross-sectional velocities, $\overline{U}$, of 0.32 and 0.40 m/s, and Reynolds numbers $Re = \bar{U}2H_{eff}/\nu$ of $2.75\cdot 10^{4}$ and $3.47\cdot 10^{4}$, respectively, where $H_{eff}$ is the distance from the surface of the PVC plates to the top wall of the channel. The regime was hydraulically smooth in all the cases investigated.
 
\begin{figure}[ht]	
	\centering
	\includegraphics[width=\columnwidth]{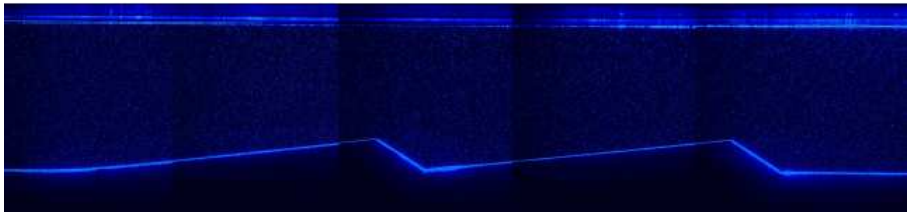}
	\caption{Example of different PIV images for the pair of ripples assembled together.}
	\label{fig:ripple_fields_together}
\end{figure}

Four and six test runs were performed for each flow rate for the tests with one single ripple and two consecutive ripples, respectively, in order to measure the flow far upstream of the ripples and over the entire ripples. PIV was used to capture the instantaneous flow fields, and all the measurements were made at the vertical symmetry plane of the channel. The light source used was a dual-cavity Nd:YAG Q-switched laser that can emit $2\times 130$ mJ at a maximum pulse rate of 15 Hz. The camera was a 7.4 $\mu$m $\times$ 7.4 $\mu$m (px$^{2}$) CCD (charge coupled device) camera with a spatial resolution of 2048 px $\times$ 2048 px that can acquire pairs of images at a maximum frequency of 10 Hz. When synchronized with the laser, the camera acquires pairs of images at a maximum frequency of 4 Hz. The total fields used measured between 60 mm $\times$ 60 mm and 70 mm $\times$ 70 mm, corresponding to a magnification of approximately 0.1. For the computations, we used cross-correlations with an interrogation area of 16 px $\times$ 16 px and $50\%$ overlap for the 70 mm $\times$ 70 mm field, corresponding to 256$^2$ interrogation areas and to a spatial resolution of 0.27 mm $\times$ 0.27 mm, and of 32 px $\times$ 32 px and $50\%$ overlap for the 60 mm $\times$ 60 mm field, corresponding to 128$^2$ interrogation areas and to a spatial resolution of 0.47 mm $\times$ 0.47 mm. Fig. \ref{fig:ripple_fields_together} shows an example of different PIV images for the pair of ripples assembled together.

Hollow glass beads 10 $\mu$m in diameter with a specific gravity of 1.05 were used as seeding particles. With these particles the laser had to be set to $66-68 \%$ of its maximum power to ensure a good balance between image contrast and undesirable reflections from the channel and ripple walls. In each of the test runs, 2000 and 3000 pairs of images were acquired for the single ripple and two ripples cases, respectively. After the experiments, the images were cross-correlated by the PIV controller software, and the instantaneous velocity fields were determined. Then, using Matlab scripts written in the course of this work, the instantaneous fields were time-averaged, and the fluctuation fields were computed and time-averaged. In the case of channel flows (unperturbed flows), the time-averaged fields were spatially averaged in the longitudinal direction. The spatial averaging was justified in that case by the fact that the flow was fully developed. For flows close to the ripple and over it (perturbed flows), the coordinates were transformed to ripple coordinates close to the bottom wall when applicable.

\section{Results}
 
\subsection{Unperturbed flow}
 
The flow upstream of the ripples was measured at a distance of more than one ripple length from the leading edge of the single or upstream ripples for both flow rates. The flow data therefore correspond to a turbulent, fully developed channel flow.

\begin{figure}[ht]
 	\centering
 	\includegraphics[width=0.7\columnwidth]{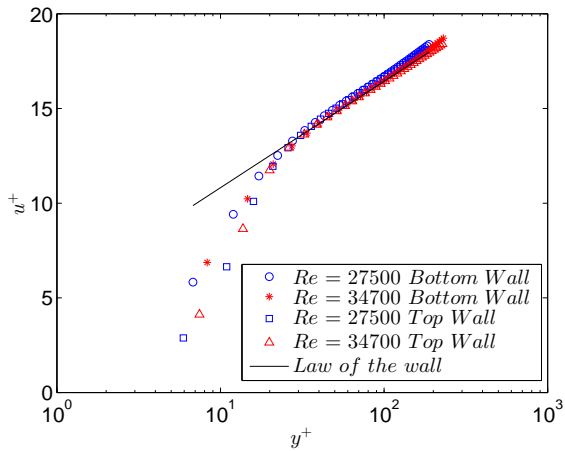}
 	\caption{Mean velocity profiles.}
 	\label{loglog}
 \end{figure}

Fig. \ref{loglog} shows the mean velocity profiles for the two Reynolds numbers. Each profile was split into two parts: one from the bottom wall to the channel center, and the other from the top wall to the channel center. The profiles are plotted in log-linear coordinates: the vertical distance from the walls normalized by the viscous length, $y^+ = yu_{*,0}/ \nu$, is on the horizontal axis on a logarithmic scale, and the mean velocity normalized by the shear velocity, $u^+ = u / u_{*,0}$, is on the vertical axis on a linear scale. The graph was plotted iteratively because the shear velocity $u_{*,0}$ was determined by fitting the experimental data in the logarithmic region ($70 < y^+ < 200$). The symbols used in the figure are explained in the key and in Tab. \ref{tab.1}.

Tab. \ref{tab.1} shows the shear velocity of the unperturbed flow $u_{*,0}$, the law-of-the-wall constant for the unperturbed flow $B_{0}$, the experimentally obtained Darcy friction factor $f_{0}$ and the Darcy friction factor from the Blasius correlation $f_{bla} = 0.316 Re_{dh} ^{-1/4}$, where $Re_{dh}$ is the Reynolds number based on the hydraulic diameter. The measurements showed that the law of the wall and Blasius correlation are valid for the turbulent flow in the test section upstream of the ripple. The flow in this region therefore corresponds to a fully developed turbulent channel flow in a hydraulic smooth regime, and the Blasius correlation can be used to estimate the unperturbed flow.

 \begin{table}[ht]
 	\begin{center}
 		\caption{Shear velocity $u_{*,0}$, law-of-the-wall constant for the unperturbed flow $B_{0}$, experimentally determined Darcy friction factor $f_{0}$ and Darcy friction factor from the Blasius correlation $f_{bla}$ for each water flow rate $Q$.}
 		\begin{tabular}{c c c c c c c c}
 			\hline\hline
 			$Q$ & $\bar{U}$ & \textit{Re} & \textit{Symbol} & $B_{0}$ & $u_{*,0}$ & $f_{0}$ & $f_{bla}$\\			
			$m^{3}/h$ & $m/s$ & $\cdots$ & $\cdots$ & $\cdots$ & $m/s$ & $\cdots$ & $\cdots$\\
 			\hline
 			8 & 0.32 & 2.75 $\times$10$^{4}$  & $\bigcirc$ & 5.41  & 0.0193 & 0.0268 & 0.0260\\
 			
 			8 & 0.32 & 2.75 $\times$10$^{4}$  & $\square$ & 5.93  & 0.0182 & 0.0237 & 0.0260\\
 			
 			10 & 0.40 & 3.47 $\times$10$^{4}$  & $\ast$ & 5.16  & 0.0237 & 0.0250 & 0.0246\\
 			
 			10 & 0.40 & 3.47 $\times$10$^{4}$  & $\triangle$ & 5.75  & 0.0227 & 0.0231 & 0.0246\\
 			\hline
 		\end{tabular}
 		\label{tab.1}
 	\end{center}
 \end{table}

\subsection{Perturbed flow}

The flow over either one or two ripples was measured for both flow rates. For comparisons between the unperturbed and perturbed flows close to the bottom wall or the ripple surface (inner region), the best approach is to use the displaced vertical coordinate $y_d = y-h$, where $h$ is the local ripple height. With $y_d$, velocities and stresses for the same distance from the wall can be compared directly. Far from the ripple surface, i.e., from the channel center (the core flow, or upper region) to the top wall, it is more convenient to use the $y$ coordinate. It is also interesting to represent the perturbation field close to the bottom wall as the difference between the perturbed velocity (i.e., the flow over the ripple) and the unperturbed velocity (in this case, the channel flow) in the displaced coordinate system, as proposed by Jackson and Hunt \cite{Jackson_Hunt}:

\begin{equation}
\Delta\vec{V} (x,y_d) = \vec{V}(x,y_d) - \vec{V_0}(x,y_d)
\label{eq2}
\end{equation}

\noindent where $\vec{V_0}(x,y_d)$ is the unperturbed flow field, which is homogeneous in $x$, $\vec{V}(x,y_d)$ is the perturbed flow field and $\Delta\vec{V} (x,y_d)$ is the perturbation field.

The corresponding mean and turbulent fields are shown next for both configurations and compared directly with the unperturbed fields. In the following, the longitudinal position $x = 0$ m corresponds to the ripple crest for the single ripple and to the crest of the upstream ripple for the pair of ripples; therefore, $x<0$ m upstream of the crest of the single or upstream ripples, and $x>0$ m downstream of the crest of the single or upstream ripples.

\begin{figure}[h!]
\begin{center}
	\begin{tabular}{c}
	\includegraphics[width=0.50\columnwidth]{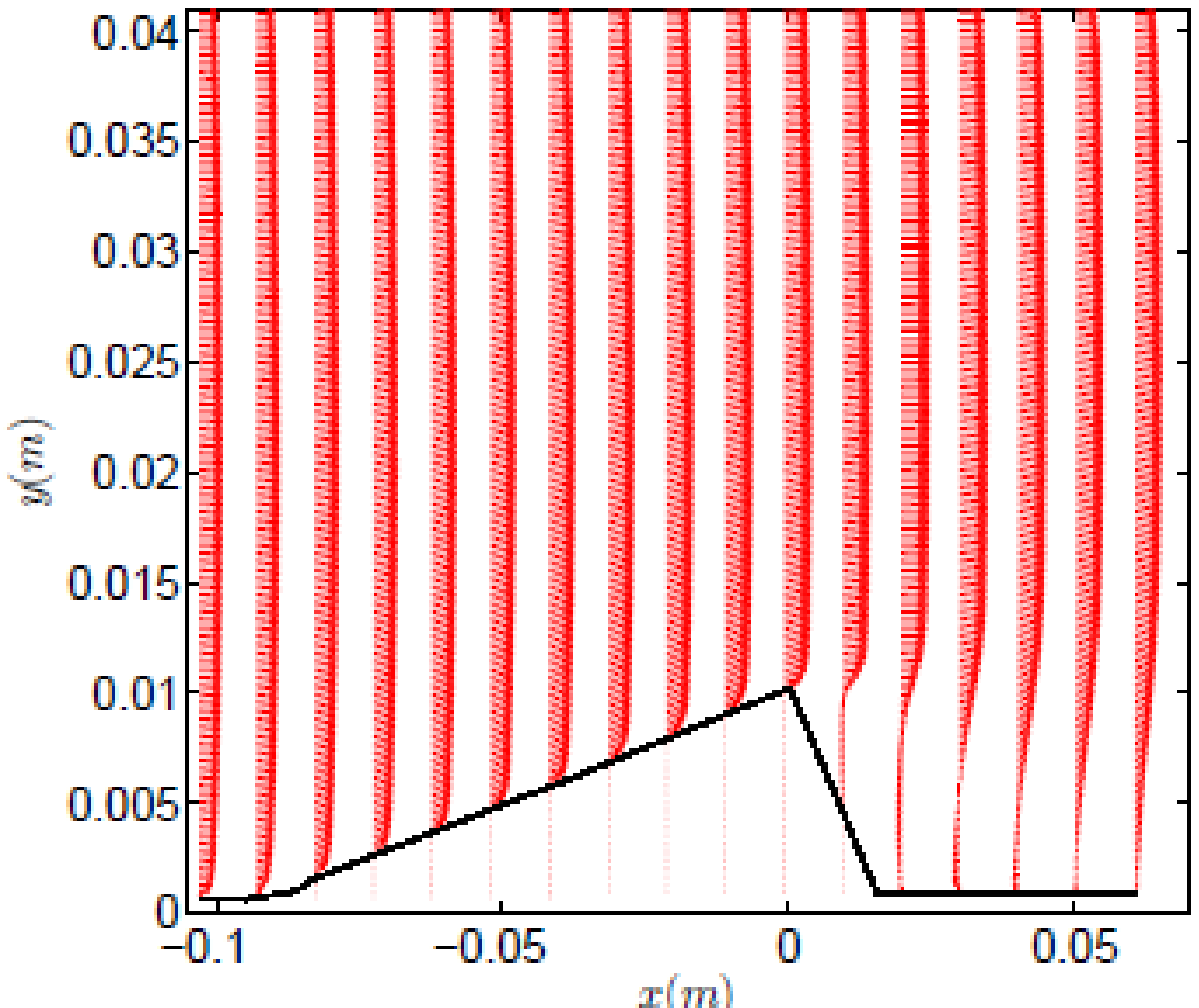}\\
	(a)\\
	\includegraphics[width=0.99\columnwidth]{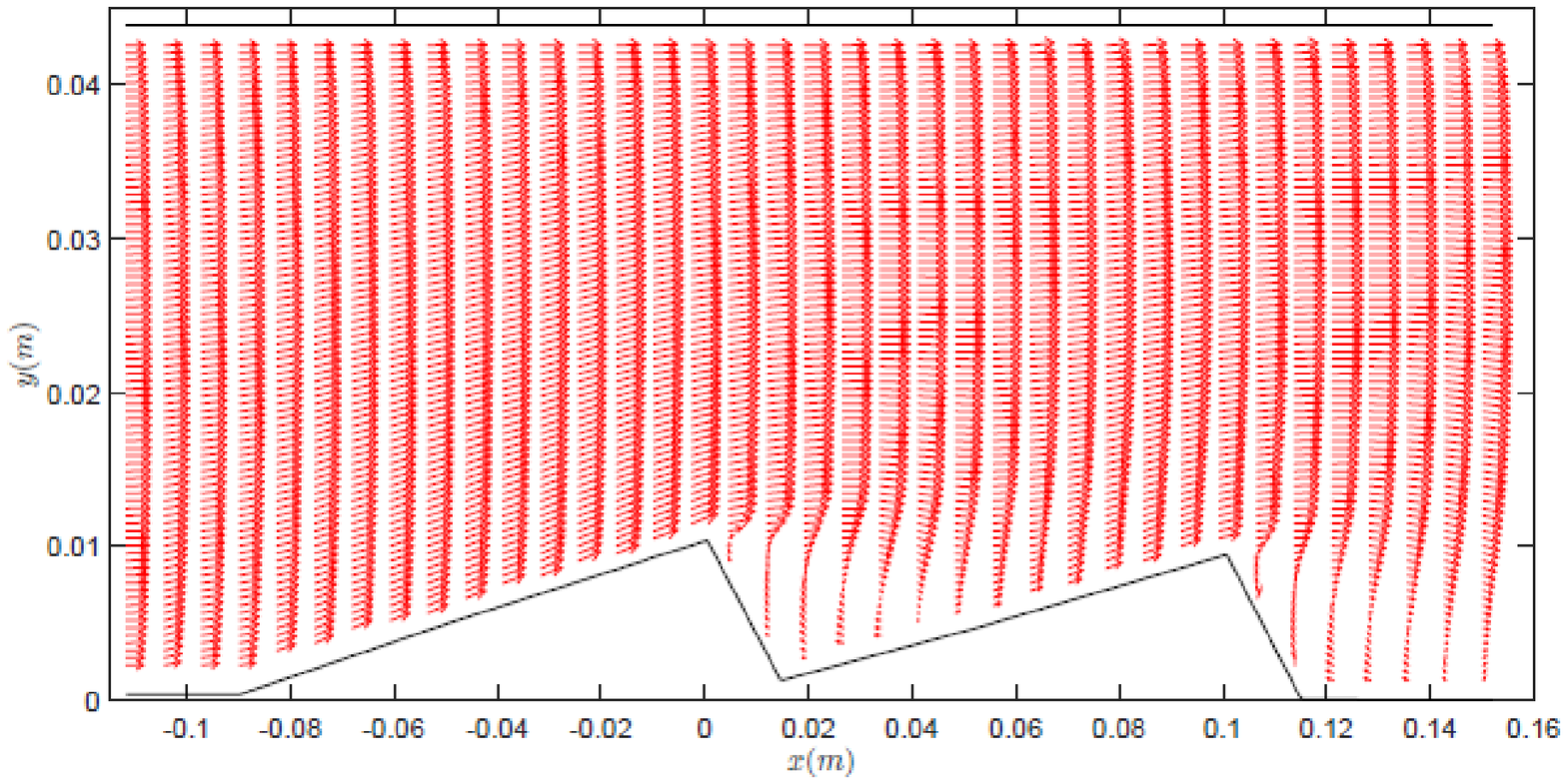}\\
	(b)
	\end{tabular} 
\end{center}
    \caption{Profiles of the mean velocities over (a) the single ripple and (b) the pair of ripples. The flow is from left to right and $Re = 2.75\cdot 10^{4}$.}
    \label{field}
\end{figure}

Fig. \ref{field} shows profiles of the mean velocity $\vec{V} (x,y) = u(x,y)\vec{i} + v(x,y)\vec{j}$ over the triangular ripples. Fig. \ref{field}a corresponds to the flow over the single ripple and Fig. \ref{field}b to the flow over the pair of ripples. The continuous line represents the surface of ripples. The main characteristics of the mean flow can be identified in Fig. \ref{field}: the flow is deviated by the ripples, and there is a recirculation bubble just downstream of the crests. For the single ripple, upstream of the crest $v$ is directed upward, and downstream of the crest it is directed upward or downward, while downstream of the ripple $u$ has positive or negative values depending on the location. For the pair of ripples, the same behavior is observed over the upstream ripple. However, over the downstream ripple the flow field is different: around 30\% of the surface of the downstream ripple lie in the recirculation region caused by the upstream ripple, with the reattachment point occurring at approximately $x$ = 0.04 m ($\approx$ 30\% of the ripple length). From this point, the flow evolves until reaching the crest, from which the flow detaches and a recirculation region similar to the one caused by the single ripple is observed. For both cases, far from the ripple surface, i.e., for higher $y$ values, $v$ is negligible. Finally, because the blockage ratio is high, $u$ is accelerated even for high $y$ values.

 \begin{figure}[ht]
 	\begin{minipage}{0.5\textwidth}
 		\begin{tabular}{c}
 			\includegraphics[width=\textwidth,clip]{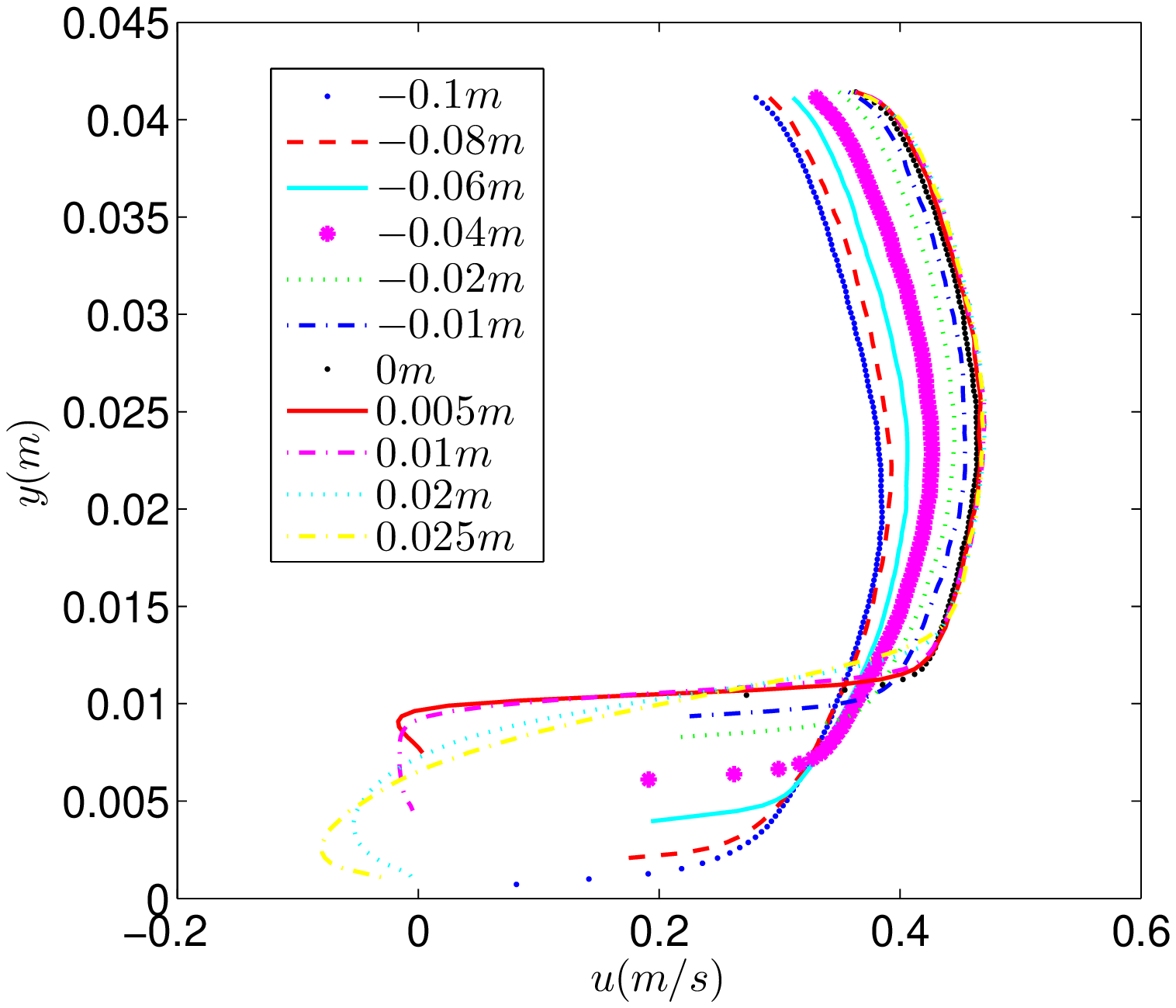}\\
 			(a)
 		\end{tabular}
 	\end{minipage}
 	\hfill
 	\begin{minipage}{0.5\textwidth}
 		\begin{tabular}{c}
 			\includegraphics[width=\textwidth,clip]{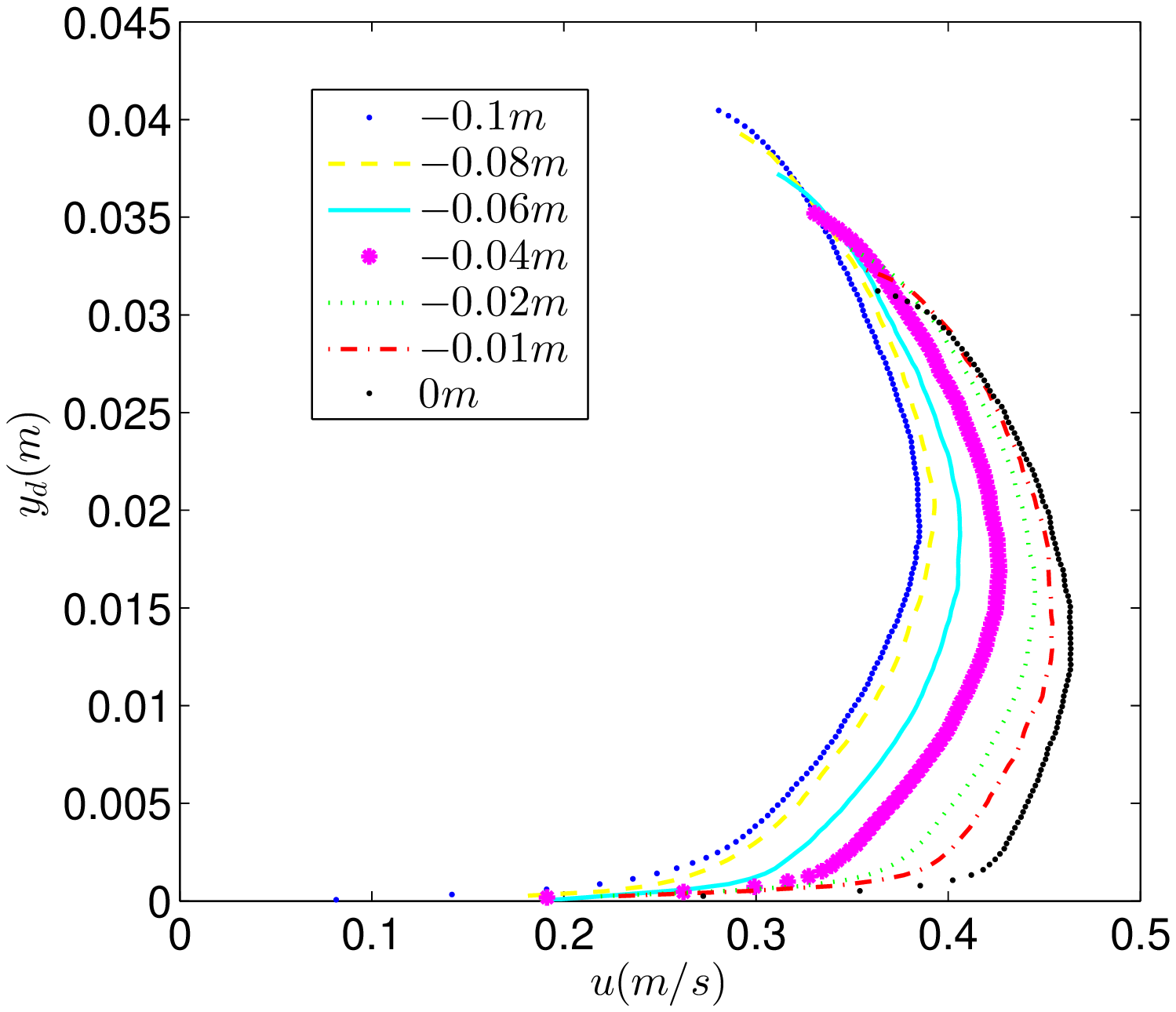}\\
 			(b)
 		\end{tabular}
 	\end{minipage}
 	\caption{(a) Longitudinal component of selected mean velocity profiles over the single ripple: (a) $u(y)$ and (b) $u(y_d)$. $Re = 2.75\cdot 10^{4}$.}
 	\label{u}
 \end{figure}

 \begin{figure}[ht]
 	\begin{minipage}{0.5\textwidth}
 		\begin{tabular}{c}
 			\includegraphics[width=\textwidth,clip]{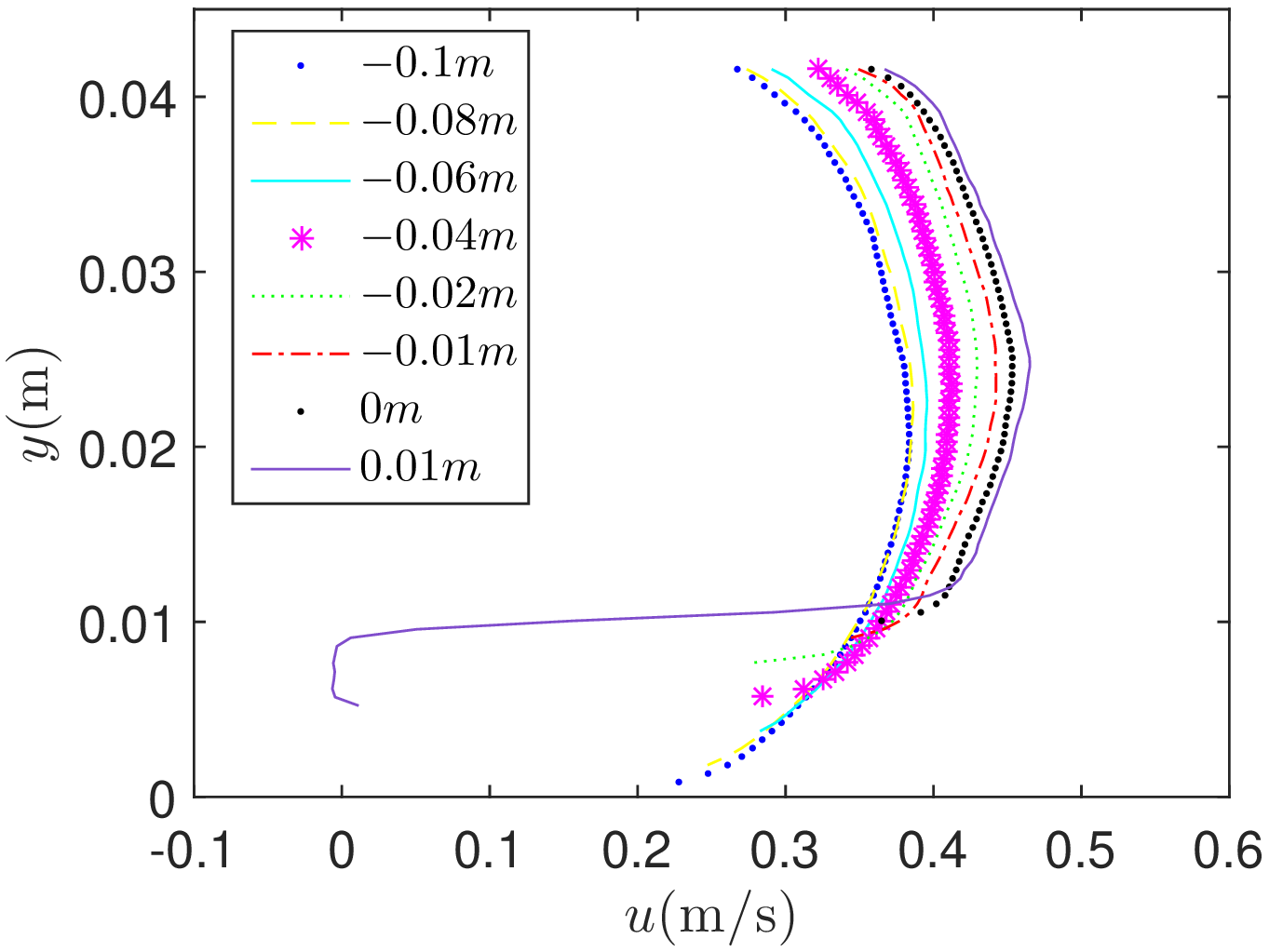}\\
 			(a)
 		\end{tabular}
 	\end{minipage}
 	\hfill
 	\begin{minipage}{0.5\textwidth}
 		\begin{tabular}{c}
 			\includegraphics[width=\textwidth,clip]{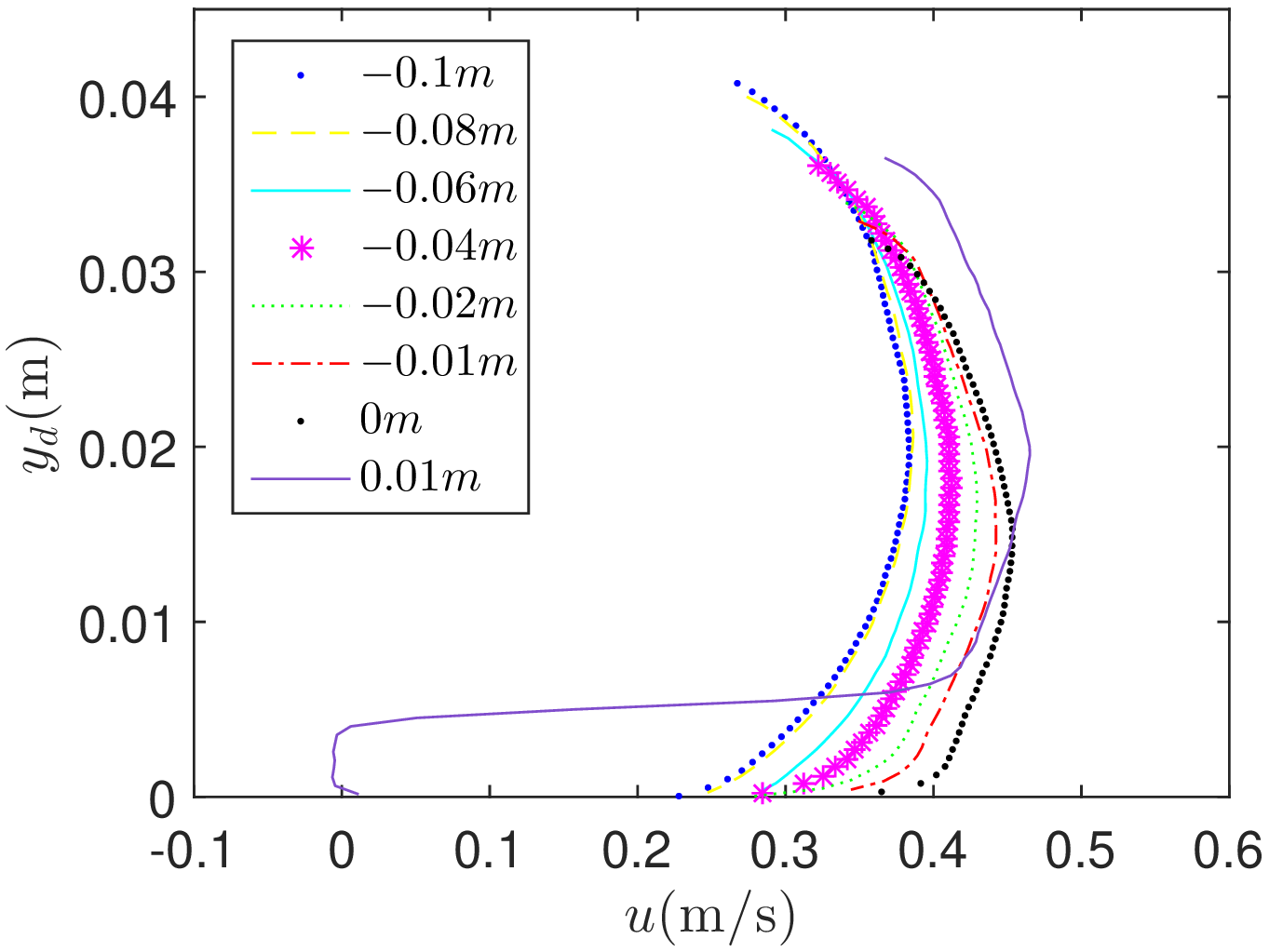}\\
 			(b)
 		\end{tabular}
 	\end{minipage}
 	\caption{Longitudinal component of selected mean velocity profiles over the upstream ripple: (a) $u(y)$ and (b) $u(y_d)$. $Re = 2.75\cdot 10^{4}$.}
 	\label{u_upstream}
 \end{figure}

 \begin{figure}[ht]
 	\begin{minipage}{0.5\textwidth}
 		\begin{tabular}{c}
 			\includegraphics[width=\textwidth,clip]{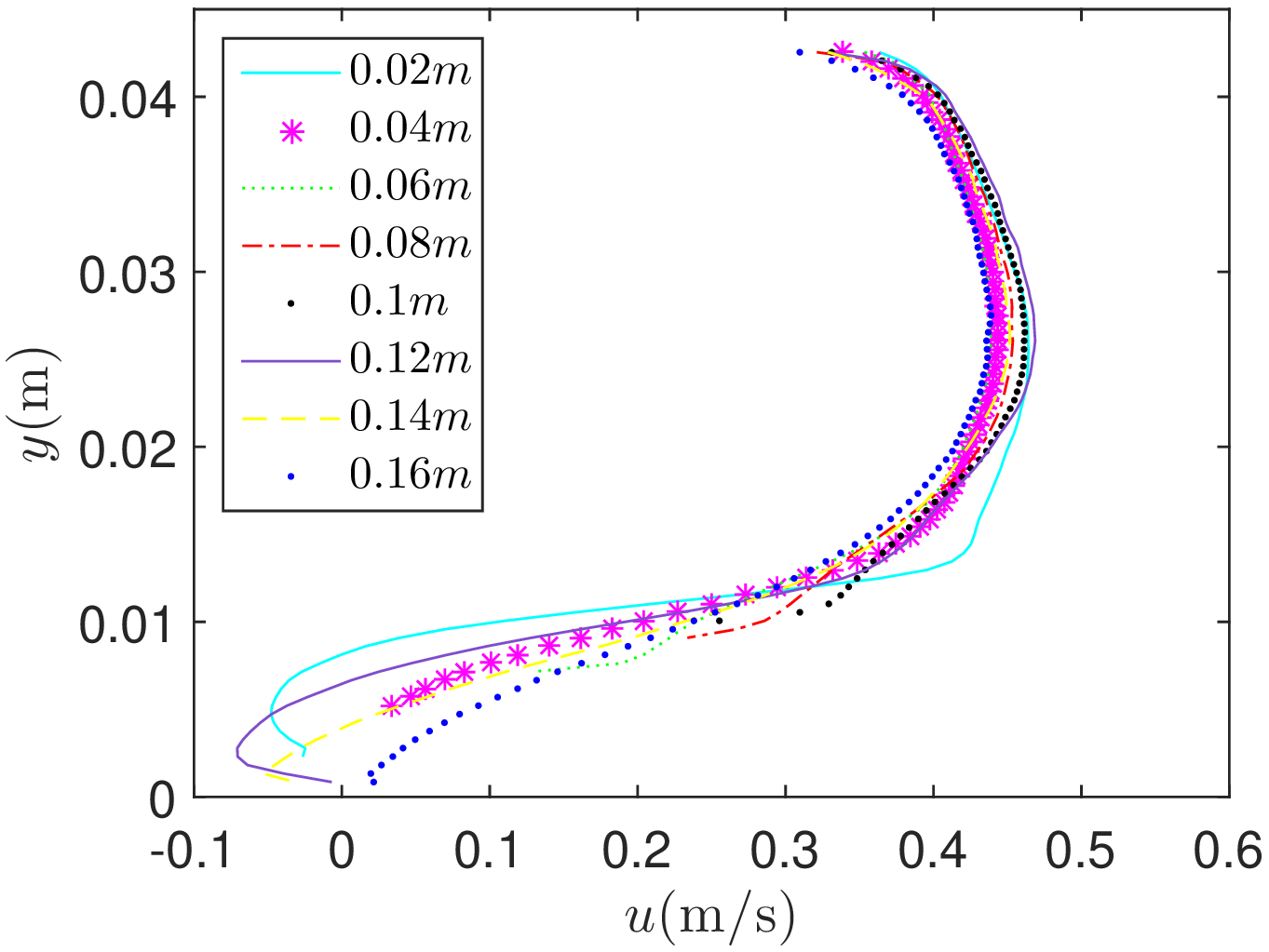}\\
 			(a)
 		\end{tabular}
 	\end{minipage}
 	\hfill
 	\begin{minipage}{0.5\textwidth}
 		\begin{tabular}{c}
 			\includegraphics[width=\textwidth,clip]{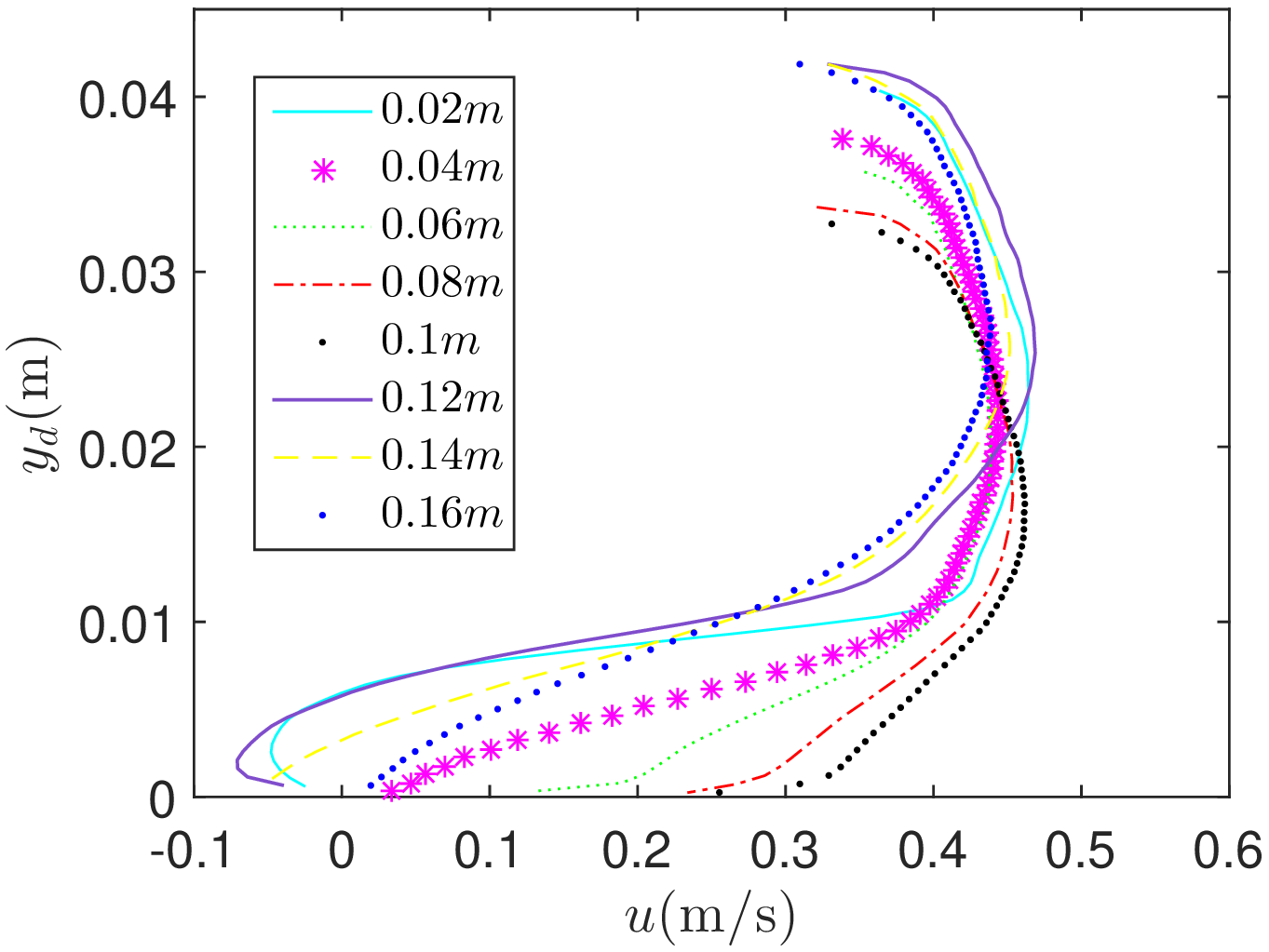}\\
 			(b)
 		\end{tabular}
 	\end{minipage}
 	\caption{Longitudinal component of selected mean velocity profiles over the downstream ripple: (a) $u(y)$ and (b) $u(y_d)$. $Re = 2.75\cdot 10^{4}$.}
 	\label{u_downstream}
 \end{figure}

Fig. \ref{u} shows the longitudinal component of selected mean velocity profiles for different longitudinal positions over the single ripple. The symbols used for different longitudinal positions are explained in the key. Fig. \ref{u}a shows $u(y)$ for longitudinal positions upstream and downstream of the ripple crest ($x=0$ m), while Fig. \ref{u}b shows $u(y_d)$ for positions upstream of the crest. Because of a strong recirculation bubble and flow confinement, the core flow is affected by the ripple and the maximum $u$ occurs around the crest, as shown in Fig. \ref{u}a. In the region $x < 0$ m, $u$ increases as the flow approaches the crest and much of the perturbation is confined to the lower region, as shown in Fig. \ref{u}b. In the region $x > 0$ m, the flow detaches and $u$ has negative values.

Figs. \ref{u_upstream} and \ref{u_downstream} show the longitudinal component of selected mean velocity profiles for the flow over the pair of ripples. Fig. \ref{u_upstream} corresponds to the flow over the upstream ripple, and the longitudinal positions are the same as in Fig. \ref{u}, while Fig. \ref{u_downstream} corresponds to the flow over the downstream ripple. Figs. \ref{u_upstream}a and \ref{u_downstream}a show $u(y)$, Figs.\ref{u_upstream}b and \ref{u_downstream}b show $u(y_d)$, and the symbols used for different longitudinal positions are explained in the key. For the upstream ripple, the same behavior described for the single ripple is observed, the main difference being the recirculation bubble, which occurs over part of the downstream ripple.

For the downstream ripple, we observe that upstream of the crest, close to the ripple leading edge, the flow is within the recirculation bubble of the upstream ripple, as shown by the $x$ = 0.02 m profile. At around $x$ = 0.04 m the flow reattaches, and downstream of this position it evolves toward the crest in a different manner from that over the upstream and single ripples. This means that the stress distribution over the downstream ripple is different from that over the upstream and single ripples. This is an important question to address because existing sand ripples are never isolated, being under velocity profiles similar to those over the downstream ripple. The stress distribution is the mechanism that sustains existing sand ripples \cite{Franklin_14}, and it is investigated in the following.

\begin{figure}[h!]
\begin{center}
	\begin{tabular}{c}
	\includegraphics[width=0.50\columnwidth]{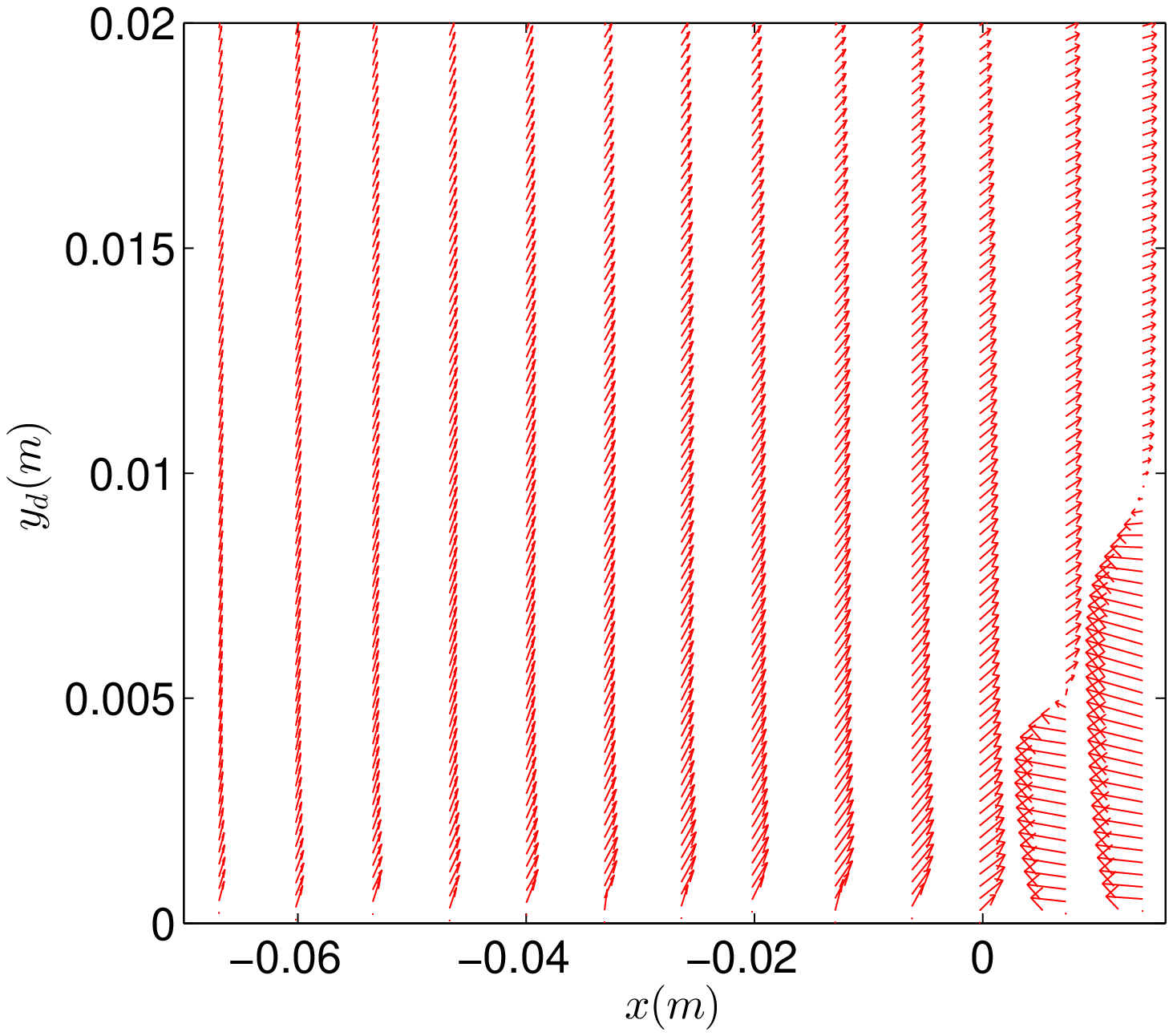}\\
	(a)\\
	\includegraphics[width=0.99\columnwidth]{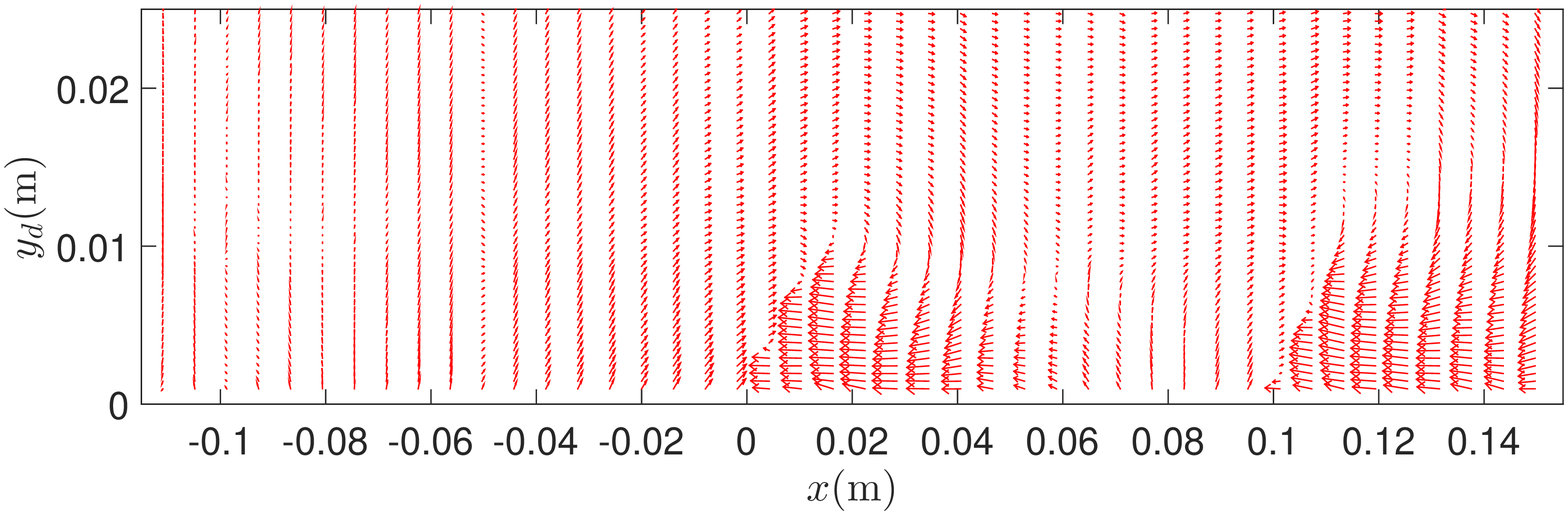}\\
	(b)
	\end{tabular} 
\end{center}
    \caption{Perturbation profiles of mean velocities for (a) the single ripple and (b) the pair of ripples. $Re = 2.75\cdot 10^{4}$.}
    \label{DV}
\end{figure}

Fig. \ref{DV} shows selected perturbation profiles of the mean velocities along the ripples. Fig. \ref{DV}a corresponds to the flow over the single ripple and Fig. \ref{DV}b to the flow over the pair of ripples. Downstream of the crests, the perturbations are negative because the flow detaches and a recirculation bubble is generated. Upstream of the crests, the perturbations increase toward the crest and are greater in the region $y_{d} < 2$ mm, which corresponds to $y_{d}^{+} < 43$, where $y_{d}^{+} = y_d u_{*,0}/ \nu $ is the displaced coordinate normalized by the viscous length of the unperturbed flow. This region corresponds to the buffer and viscous layers of the unperturbed flow, where the viscous stresses are significant \cite{Schlichting_1}.

 \begin{figure}[ht]
 	\begin{minipage}{0.5\textwidth}
 		\begin{tabular}{c}
 			\includegraphics[width=\textwidth,clip]{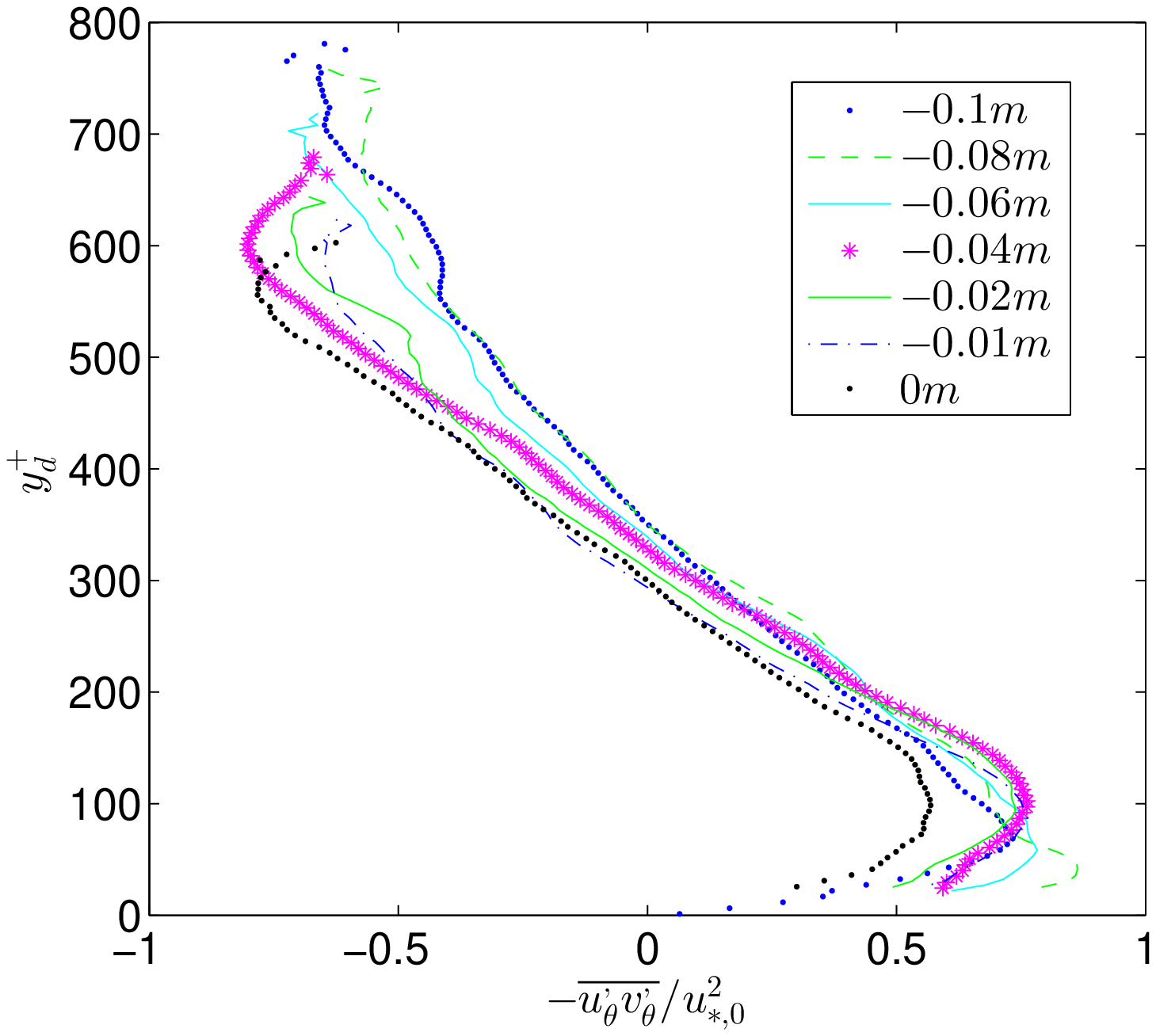}\\
 			(a)
 		\end{tabular}
 	\end{minipage}
 	\hfill
 	\begin{minipage}{0.5\textwidth}
 		\begin{tabular}{c}
 			\includegraphics[width=\textwidth,clip]{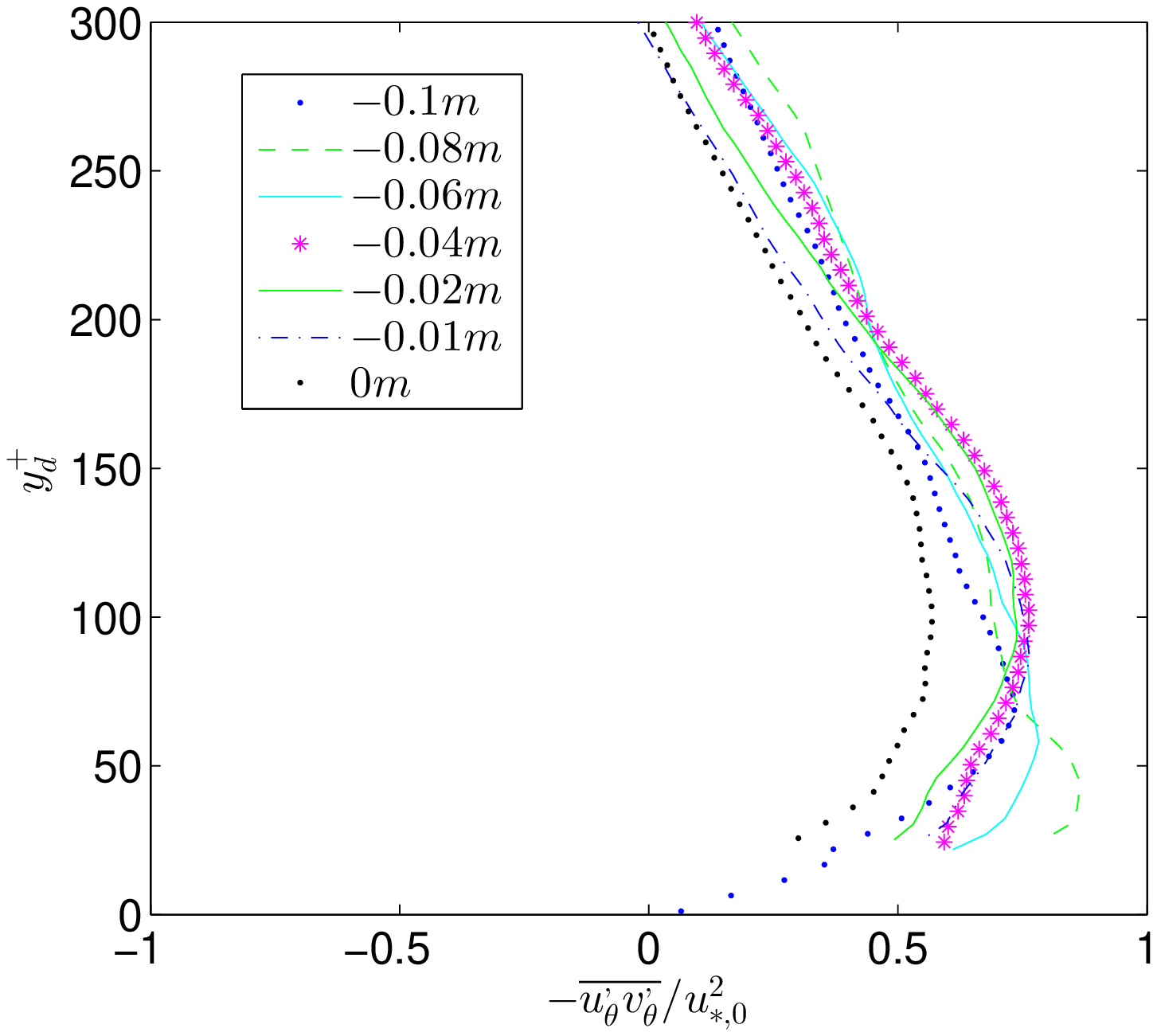}\\
 			(b)
 		\end{tabular}
 	\end{minipage}
 	\caption{Profiles of the $xy$ component of the Reynolds stress aligned with the ripple surface in dimensionless form for the single ripple and $Re = 2.75\cdot 10^{4}$. (a) Profile over the entire channel height. (b) Detail of the profile in the region $y_d^+ \leq 300$.}
 	\label{recp}
 \end{figure}

Within the viscous layer, viscous stresses are significant, while turbulent stresses are negligible. Just above this layer, in the buffer layer, viscous and turbulent stresses are significant, and well above the viscous layer, within the overlap layer, turbulent stresses are significant and viscous stresses negligible. In the case of local-equilibrium buffer and viscous layers, turbulent stresses in the overlap layer are related to viscous stresses in the viscous layer and are of the same order of magnitude. It is for this reason that, in the case of boundary layers with local-equilibrium inner regions, the wall shear stress can be computed from the shear velocity, which scales with the square root of turbulent stresses. Investigation of turbulent and viscous stresses can indicate if the flow is in local equilibrium in inner regions.

 \begin{figure}[ht]
 	\begin{minipage}{0.5\textwidth}
 		\begin{tabular}{c}
 			\includegraphics[width=\textwidth,clip]{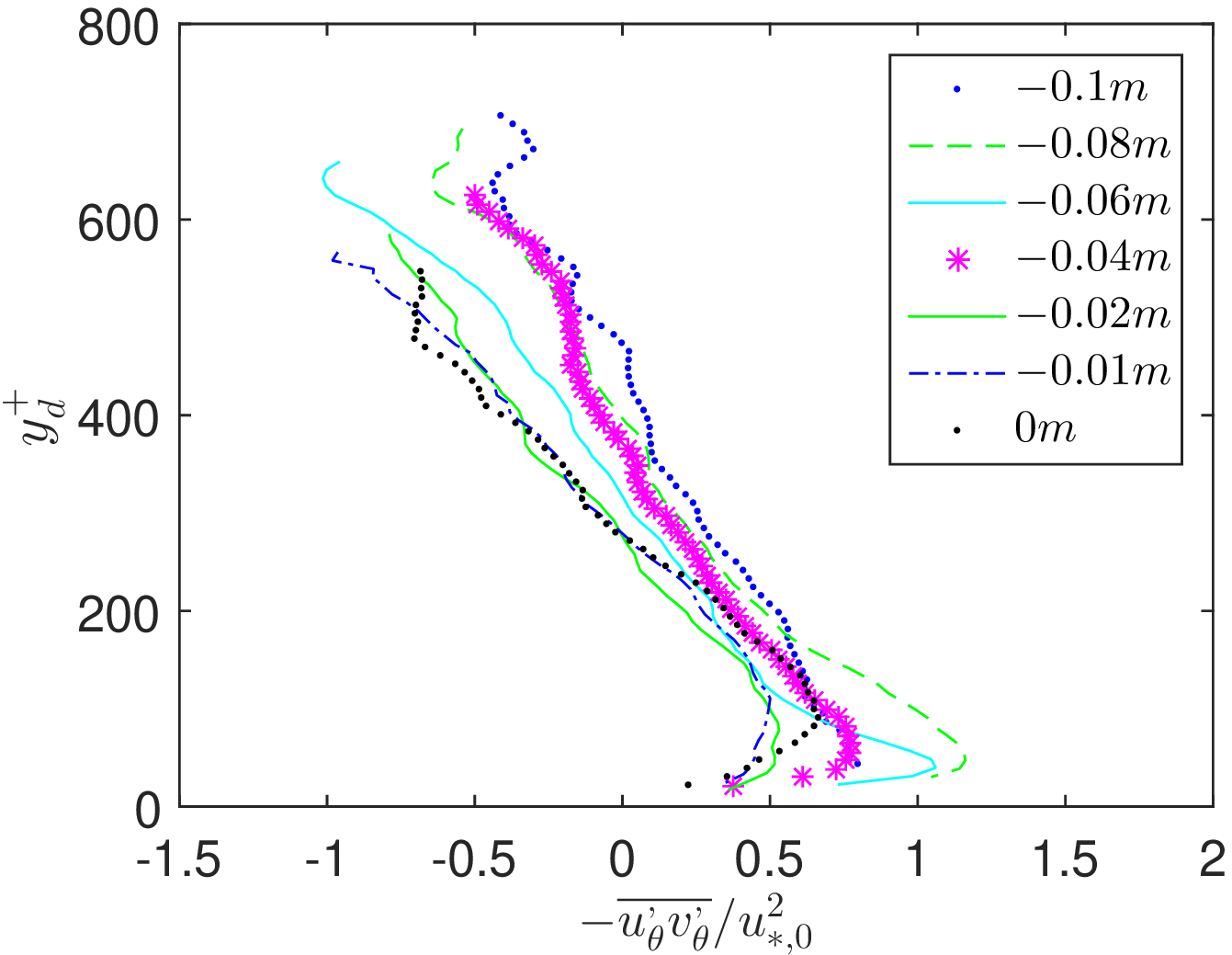}\\
 			(a)
 		\end{tabular}
 	\end{minipage}
 	\hfill
 	\begin{minipage}{0.5\textwidth}
 		\begin{tabular}{c}
 			\includegraphics[width=\textwidth,clip]{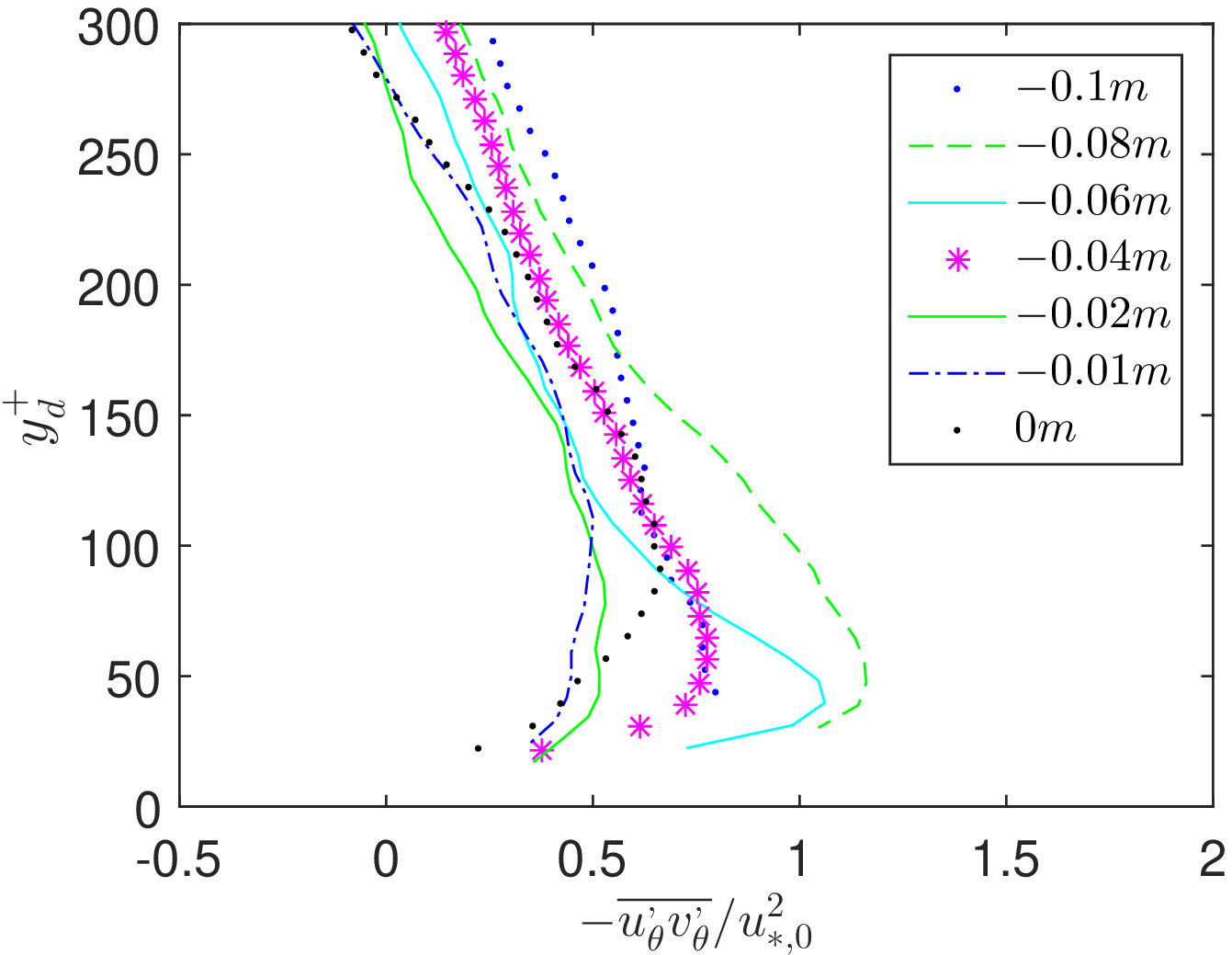}\\
 			(b)
 		\end{tabular}
 	\end{minipage}
 	\caption{Profiles of the $xy$ component of the Reynolds stress aligned with the ripple surface in dimensionless form for the upstream ripple and $Re = 2.75\cdot 10^{4}$. (a) Profile over the entire channel height. (b) Detail of the profile in the region $y_d^+ \leq 300$.}
 	\label{recp2}
 \end{figure}

 \begin{figure}[ht]
 	\begin{minipage}{0.5\textwidth}
 		\begin{tabular}{c}
 			\includegraphics[width=\textwidth,clip]{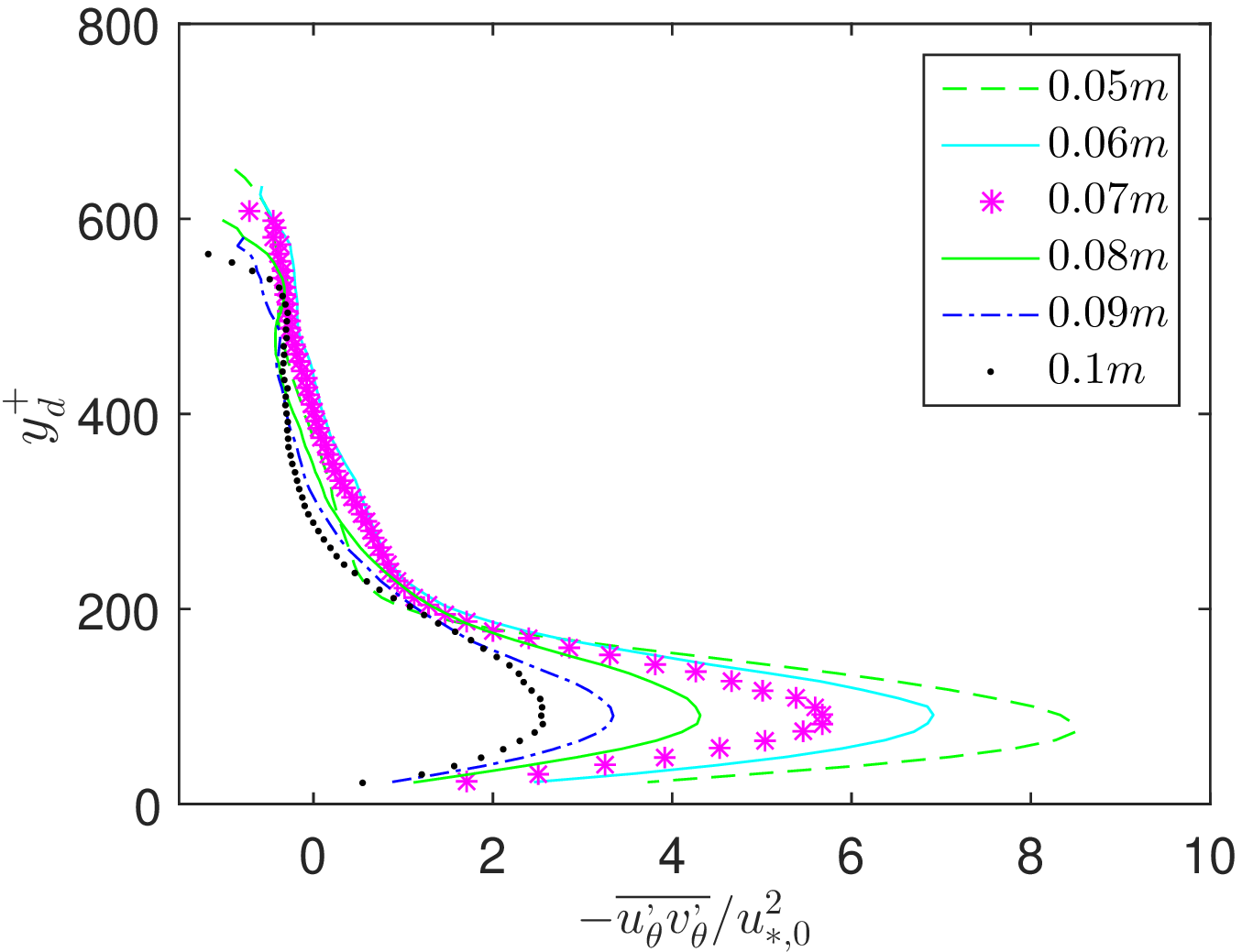}\\
 			(a)
 		\end{tabular}
 	\end{minipage}
 	\hfill
 	\begin{minipage}{0.5\textwidth}
 		\begin{tabular}{c}
 			\includegraphics[width=\textwidth,clip]{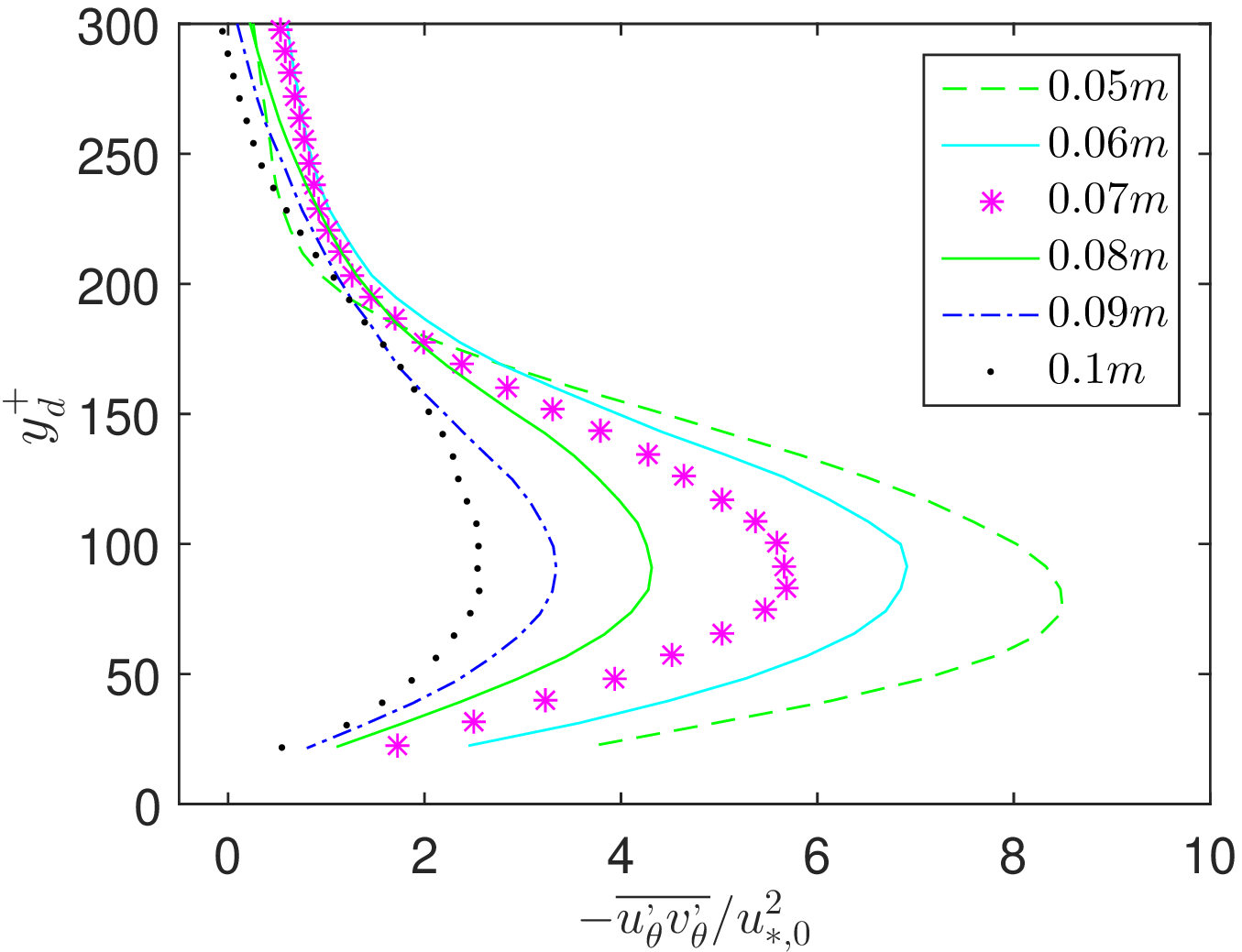}\\
 			(b)
 		\end{tabular}
 	\end{minipage}
 	\caption{Profiles of the $xy$ component of the Reynolds stress aligned with the ripple surface in dimensionless form for the downstream ripple and $Re = 2.75\cdot 10^{4}$. (a) Profile over the entire channel height. (b) Detail of the profile in the region $y_d^+ \leq 300$.}
 	\label{recp3}
 \end{figure}

Figs. \ref{recp} to \ref{recp3} show selected profiles of the $xy$ component of the Reynolds stress aligned with the ripple surface in dimensionless form, $-\overline{u'_{\theta} v'_{\theta}} / u_{*,0}^{2}$, in the regions upstream of the ripple crests. The $-\overline{u'_{\theta} v'_{\theta}}$ profiles were averaged by a sliding-window process over the closest nine points to reduce noise. Figs. \ref{recp}, \ref{recp2} and \ref{recp3} correspond to the flow over the single, upstream and downstream ripples, respectively. The behaviors of the single and upstream ripples are similar. They are compatible with a local-equilibrium inner region because the values of $-\overline{u'_{\theta} v'_{\theta}}$ in the region $20 < y_d^{+} < 250$ scale with $\tau_{0}/\rho = u_{*,0}^2$, which in turn, when perturbed, becomes $\tau/ \rho = u_*^2$. Therefore, $-\overline{u'_{\theta} v'_{\theta}}$ and $u_{*,0}^{2}$ are of the same order. In addition, considering each individual profile in the region $30 < y^+ < 200$, the values of $-\overline{u'_{\theta} v'_{\theta}}$ reach a maximum at $y^+ \approx 100$ and decay toward the wall, indicating that viscous stresses increase near the wall. In this case, logarithmic profiles for the mean velocity can be expected in the overlap regions over the ripple upstream of the crest. However, at the crest ($x=0$ m), the maximum turbulent stress is much smaller than at other positions, which may indicate a deviation from local equilibrium. We note, finally, that $-\overline{u'_{\theta} v'_{\theta}}$ increases and then decreases toward the crest.

From Fig. \ref{recp3}, which presents only the profiles downstream of the reattachment point, we observe that the behavior is different for the downstream ripple. The maxima of the $-\overline{u'_{\theta} v'_{\theta}} / u_{*,0}^{2}$ profiles show stronger values than over the upstream and single ripples, and those values decrease toward the ripple crest. Far upstream of the crest, the $xy$ component of the Reynolds stress reaches values greater than 8 times that of the unperturbed shear stress, and at the crest it has still more than 2 times the unperturbed value. These high values are caused by the recirculation bubble and the flow reattachment on the surface of the downstream ripple. The reattached flow is completely different from the channel flow reaching the upstream and single ripples, so that it evolves toward the ripple crest in a different way when compared to the flow over the other ripples. Considering each individual profile, they show large peaks at $y^+ \approx 100$; therefore, the existence of a region of constant stress within $y_d^+ < 100$, and with logarithmic profiles for the mean velocity, is not expected over the downstream ripple (as we show next). 

To investigate the total shear stress, the viscous shear stress along the ripples must be calculated. This was done using the expression:

\begin{equation}
\tau_{visc} \approx \mu\frac{\partial u_{\theta}}{\partial y_{d,\theta}}
\label{eq3}
\end{equation}

\noindent where $u_{\theta}$ is the component of the mean velocity aligned with the surface of the triangular ripple, and $y_{d,\theta}$ is a displaced coordinate perpendicular to the ripple surface.

 \begin{figure}[ht]
 	\begin{minipage}{0.5\textwidth}
 		\begin{tabular}{c}
 			\includegraphics[width=\textwidth,clip]{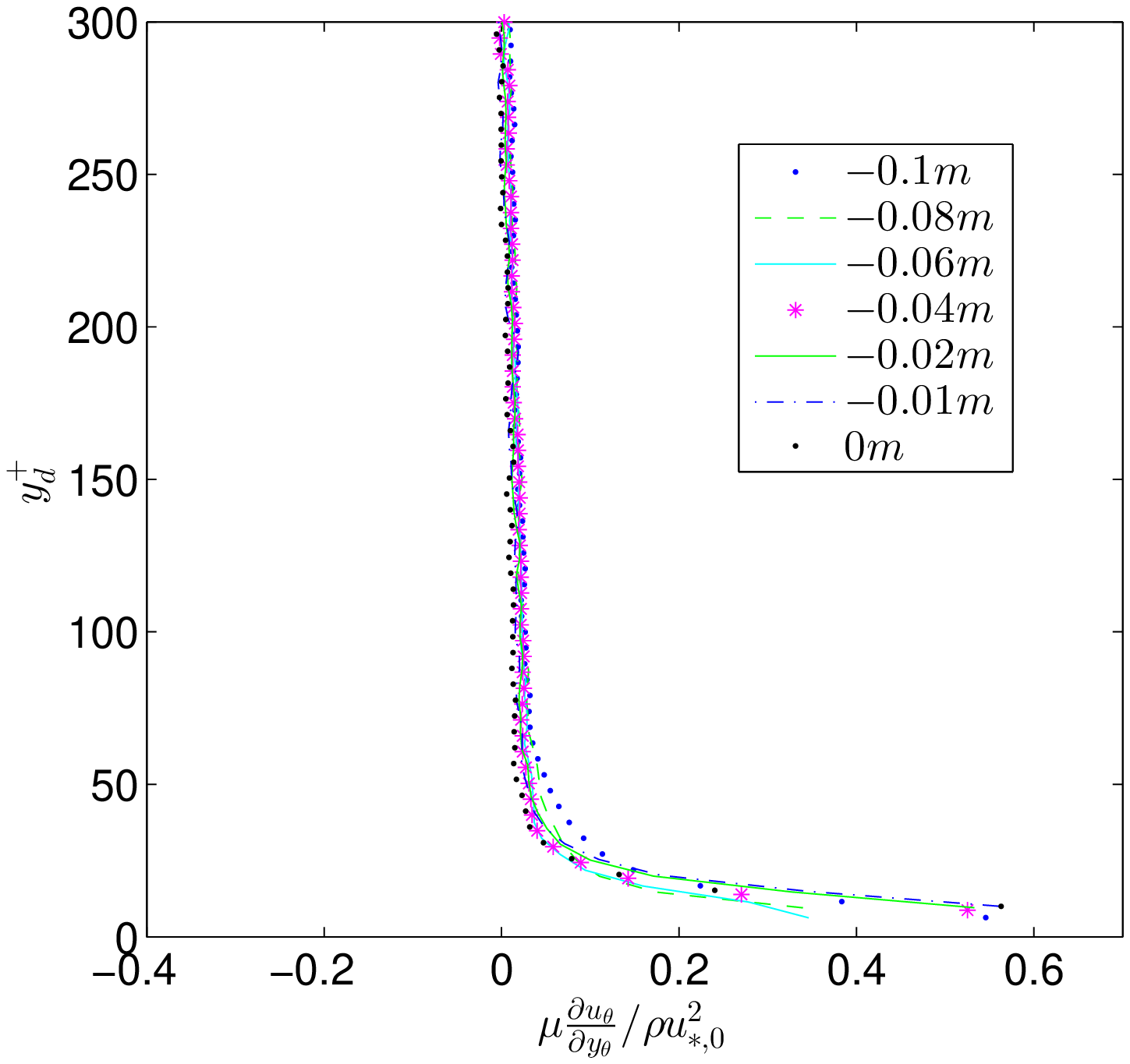}\\
 			(a)
 		\end{tabular}
 	\end{minipage}
 	\hfill
 	\begin{minipage}{0.5\textwidth}
 		\begin{tabular}{c}
 			\includegraphics[width=\textwidth,clip]{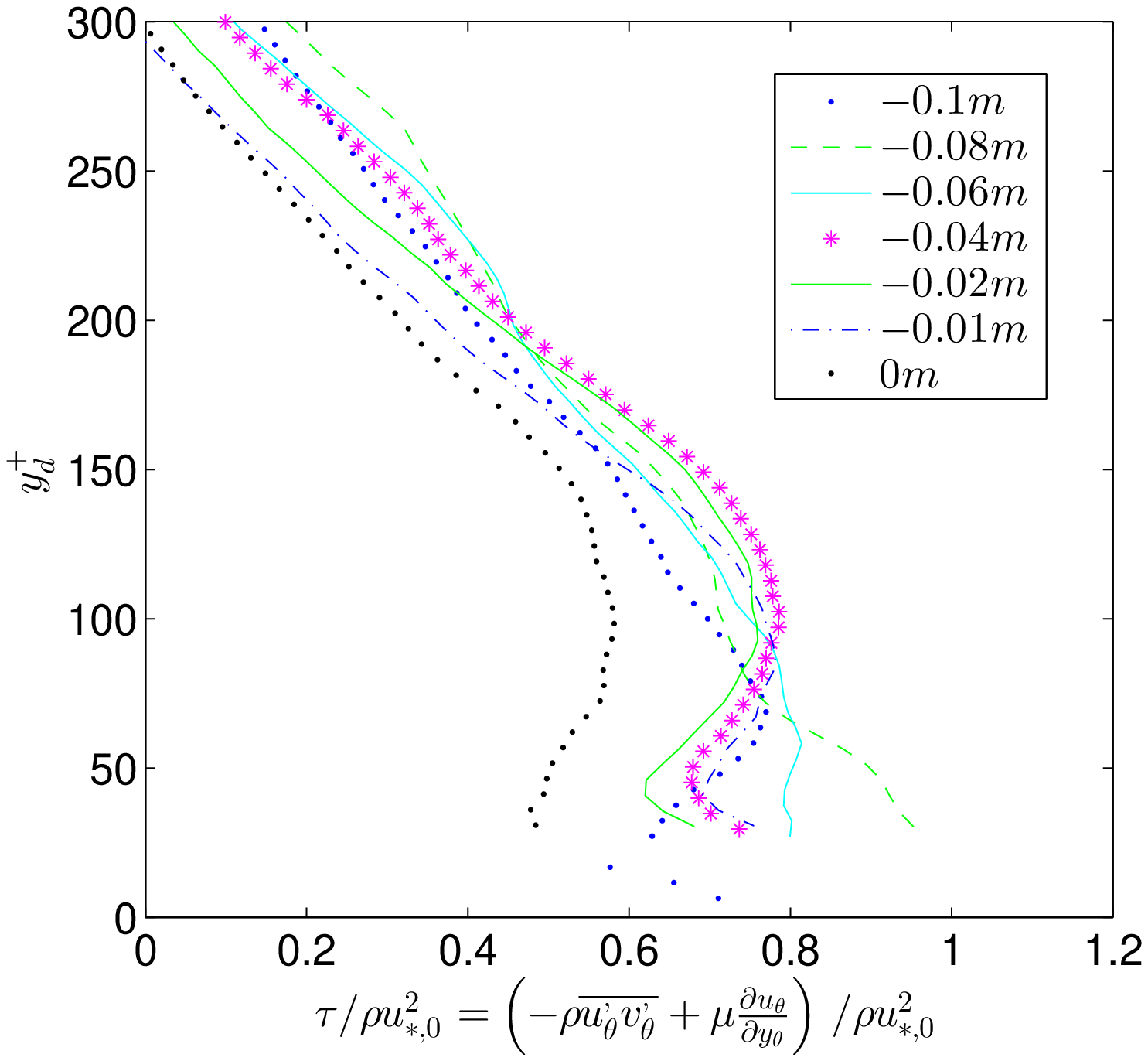}\\
 			(b)
 		\end{tabular}
 	\end{minipage}
 	\caption{(a) Selected profiles of the $xy$ component of the viscous shear stress aligned with the ripple surface in dimensionless form and (b) profiles of the $xy$ component of the total shear stress aligned with the ripple surface in dimensionless form for the single ripple and $Re = 2.75\cdot 10^{4}$.}
 	\label{fig_total_stress}
 \end{figure}

\begin{figure}[ht]
 	\begin{minipage}{0.5\textwidth}
 		\begin{tabular}{c}
 			\includegraphics[width=\textwidth,clip]{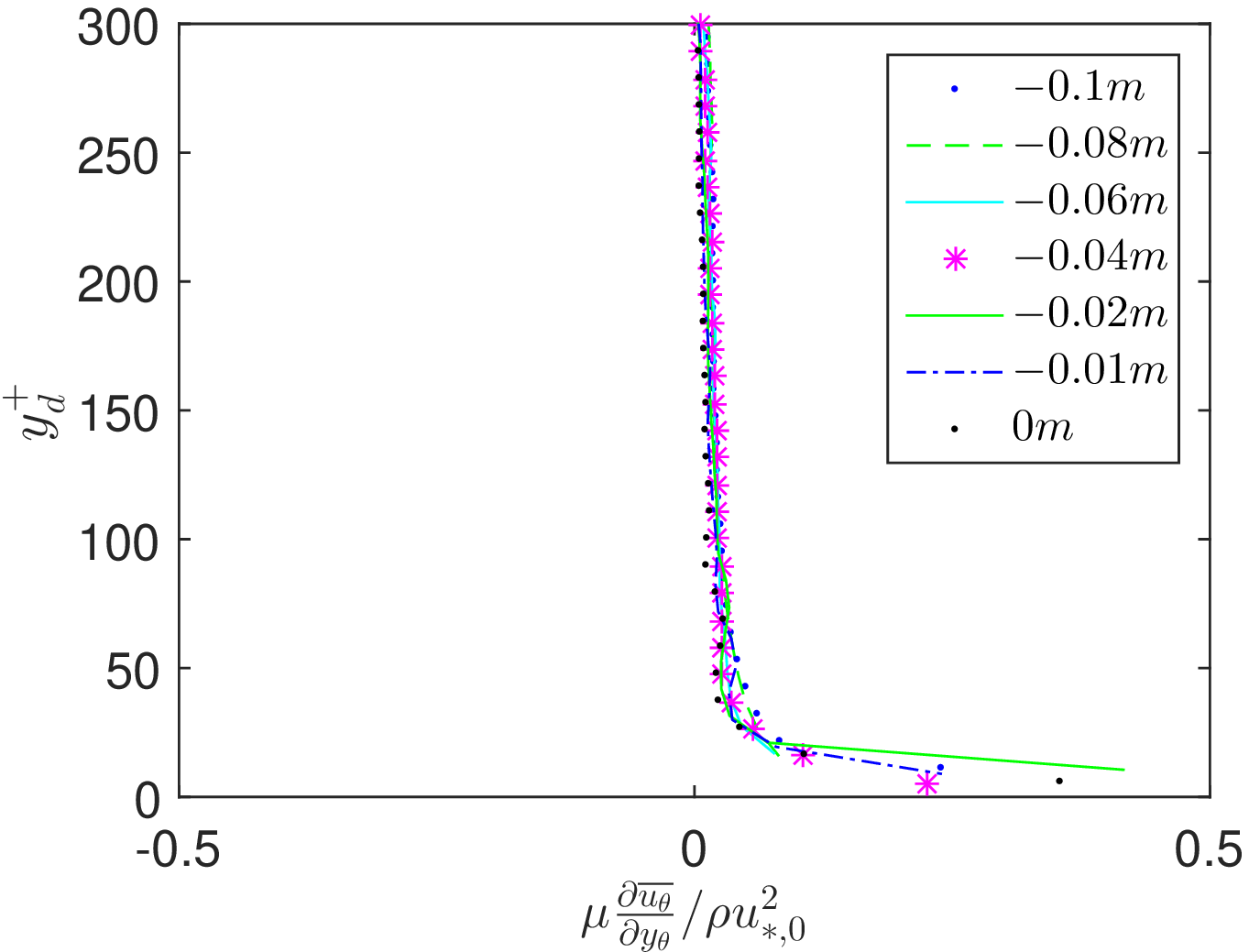}\\
 			(a)
 		\end{tabular}
 	\end{minipage}
 	\hfill
 	\begin{minipage}{0.5\textwidth}
 		\begin{tabular}{c}
 			\includegraphics[width=\textwidth,clip]{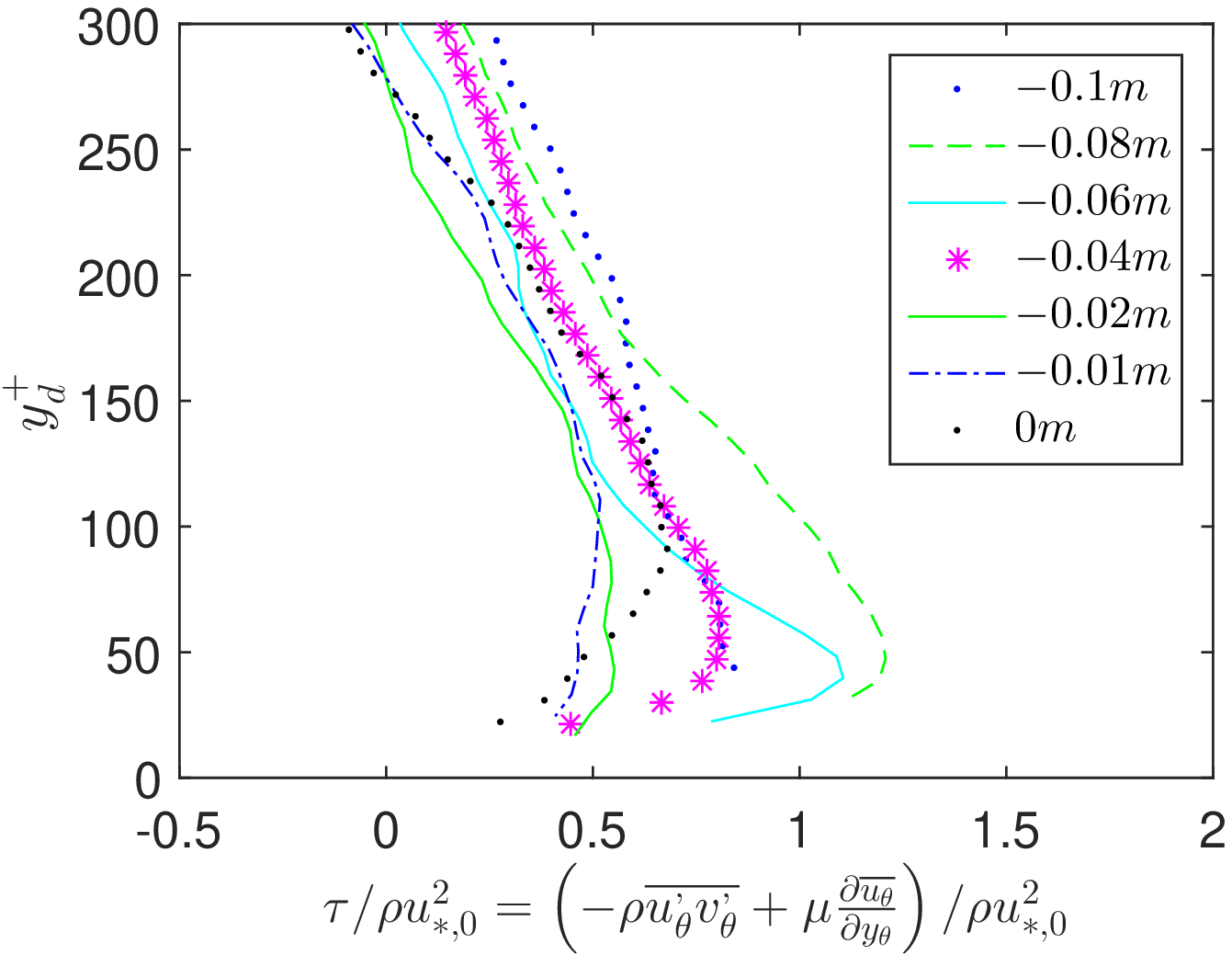}\\
 			(b)
 		\end{tabular}
 	\end{minipage}
 	\caption{(a) Selected profiles of the $xy$ component of the viscous shear stress aligned with the ripple surface in dimensionless form and (b) profiles of the $xy$ component of the total shear stress aligned with the ripple surface in dimensionless form for the upstream ripple and $Re = 2.75\cdot 10^{4}$.}
 	\label{fig_total_stress_upstream}
 \end{figure}

\begin{figure}[ht]
 	\begin{minipage}{0.5\textwidth}
 		\begin{tabular}{c}
 			\includegraphics[width=\textwidth,clip]{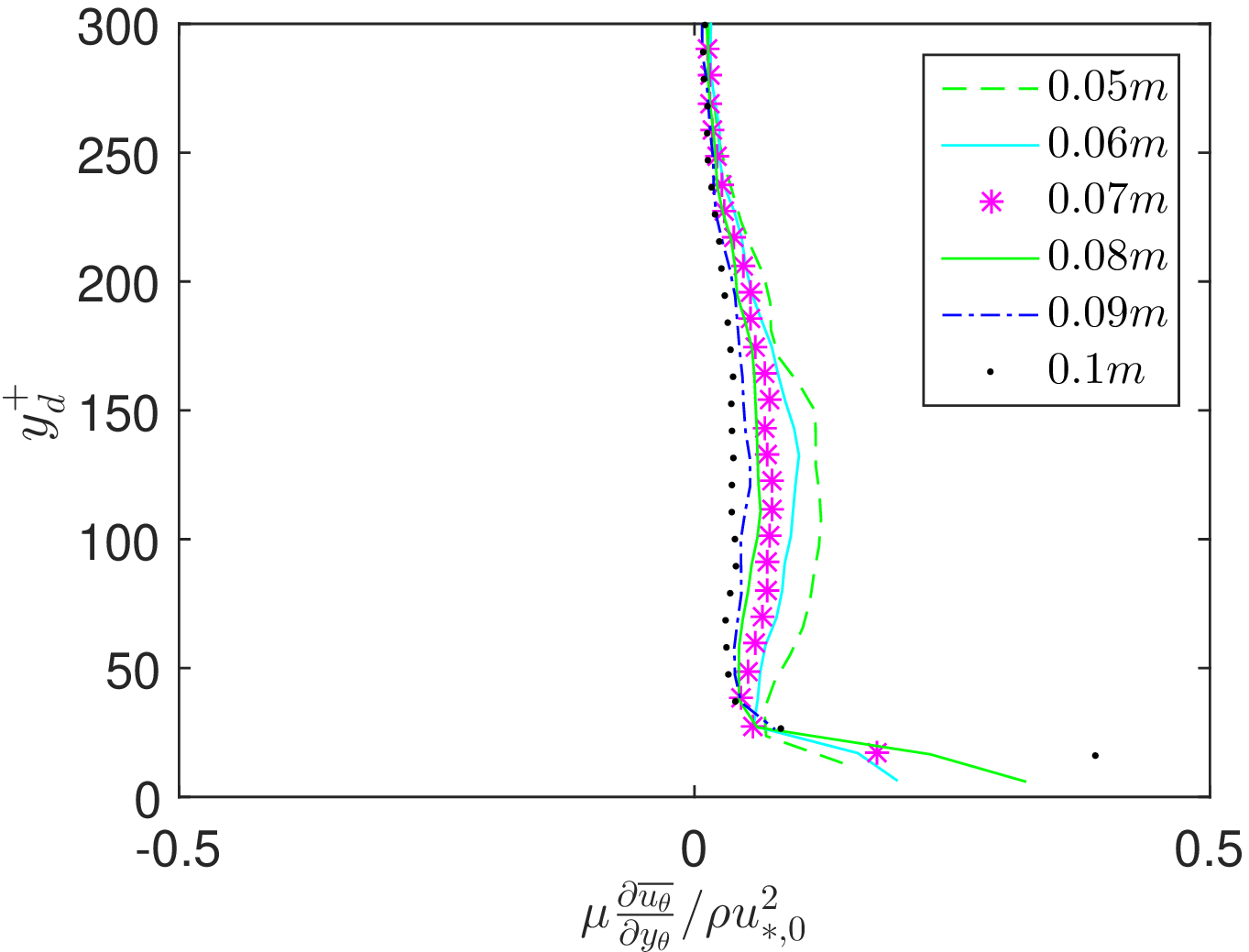}\\
 			(a)
 		\end{tabular}
 	\end{minipage}
 	\hfill
 	\begin{minipage}{0.5\textwidth}
 		\begin{tabular}{c}
 			\includegraphics[width=\textwidth,clip]{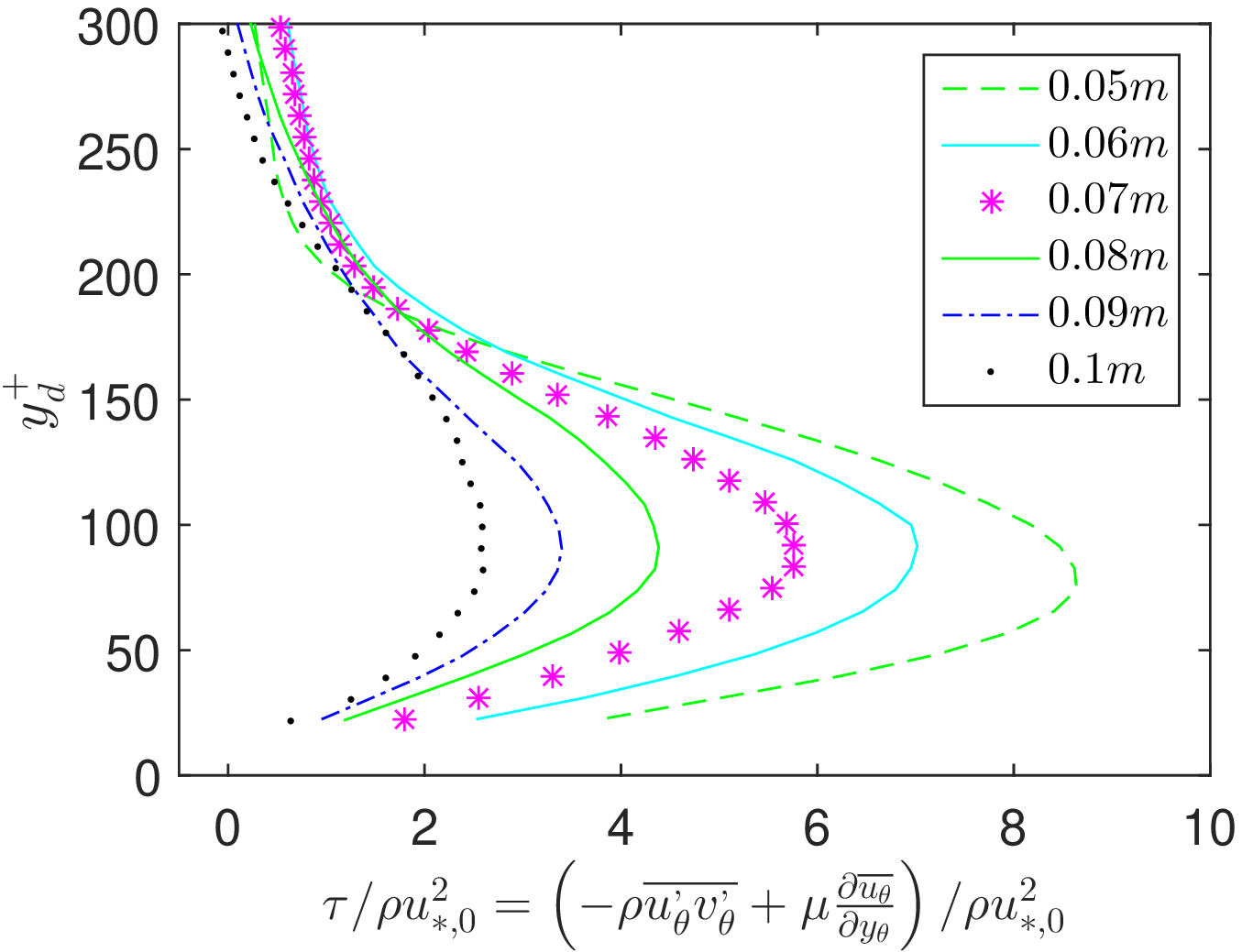}\\
 			(b)
 		\end{tabular}
 	\end{minipage}
 	\caption{(a) Selected profiles of the $xy$ component of the viscous shear stress aligned with the ripple surface in dimensionless form and (b) profiles of the $xy$ component of the total shear stress aligned with the ripple surface in dimensionless form for the downstream ripple and $Re = 2.75\cdot 10^{4}$.}
 	\label{fig_total_stress_downstream}
 \end{figure}

Figs. \ref{fig_total_stress}a, \ref{fig_total_stress_upstream}a and \ref{fig_total_stress_downstream}a show selected profiles of the $xy$ component of the viscous shear stress aligned with the ripple surface in dimensionless form,  $\mu\frac{\partial u_{\theta}}{\partial y_{d,\theta}} \left( \rho u_{*,0}^2 \right) ^{-1}$, for the single, upstream and downstream ripples, respectively, and $Re = 2.75\cdot 10^{4}$. The profiles in Figs. \ref{fig_total_stress}a, \ref{fig_total_stress_upstream}a and \ref{fig_total_stress_downstream}a are at the same longitudinal positions as those in Figs. \ref{recp}, \ref{recp2} and \ref{recp3}, respectively. The profiles for the single and upstream ripples are similar, with the viscous shear stress being significant only in the $y_d^+ < 50$ region and decaying as $y_d^+$ increases from 0 to $y_d^+ = 50$ in proportion to the increase in the turbulent stress. Fig. \ref{fig_total_stress_downstream}a presents only the profiles downstream of the reattachment point. We observe that the behavior is slightly different for the downstream ripple. As for the other cases, the viscous shear stress is significant in the $y_d^+ < 50$ region and decays from $y_d^+$ = 0 to $y_d^+$ = 50; however, we observe in the individual profiles the presence of a bump in the 50 $\leq\, y_d^+\, \leq$ 200 region, that decreases toward the crest. As observed in the case of Fig. \ref{recp3}, the reattached flow is completely different from the channel flow reaching the upstream and single ripples, so that it evolves in a different way when compared to the flow over the other ripples.

\begin{sloppypar}
Figs. \ref{fig_total_stress}b, \ref{fig_total_stress_upstream}b and \ref{fig_total_stress_downstream}b show the profiles of the $xy$ component of the total shear stress aligned with the ripple surface in dimensionless form, $\left( \mu\frac{\partial u_{\theta}}{\partial y_{d,\theta}} -\rho \overline{u'_{\theta} v'_{\theta}} \right) \left( \rho u_{*,0}^2 \right) ^{-1}$, for the single, upstream and downstream ripples, respectively, and $Re = 2.75\cdot 10^{4}$. The profiles in Figs. \ref{fig_total_stress}b, \ref{fig_total_stress_upstream}b and \ref{fig_total_stress_downstream}b are at the same longitudinal positions as those in Figs. \ref{recp}, \ref{recp2} and \ref{recp3}, respectively. Again, the profiles for the single and upstream ripples are similar. Taking into account that the uncertainties caused by undesired reflections from the walls are higher close to the surface of ripples (in the buffer and viscous layers), we can consider that there is a region around $20 < y_d^+ < 100$ where the total stress is roughly constant, which is, again, consistent with a local-equilibrium inner region. For the downstream ripple, we do not observe the existence of a region of constant stress within $y_d^+ < 100$, so that a local-equilibrium inner region is absent. This absence of local equilibrium is caused by the flow separation (recirculation bubble and flow reattachment), which makes the flow to evolve in a different way when compared to the flow over the other ripples.
\end{sloppypar}
 
 \begin{figure}[ht]
 	\centering
 	\includegraphics[width=0.5\columnwidth]{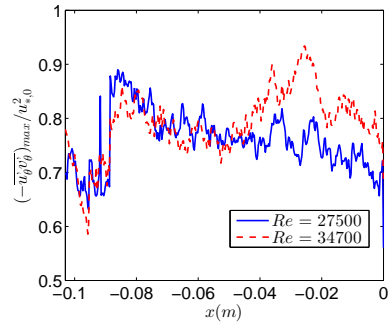}
 	\caption{Maxima of the normalized Reynolds stress $(-\overline{u'_{\theta} v'_{\theta}})_{max}/({u}_{*,0}^{2})$ as a function of longitudinal position $x$ for the single ripple. The continuous and dashed lines correspond to $Re = 2.75\times 10^{4}$ and $Re = 3.47\times 10^{4}$, respectively.}
 	\label{remax}
 \end{figure}

\begin{figure}[ht]
 	\begin{minipage}{0.5\textwidth}
 		\begin{tabular}{c}
 			\includegraphics[width=\textwidth,clip]{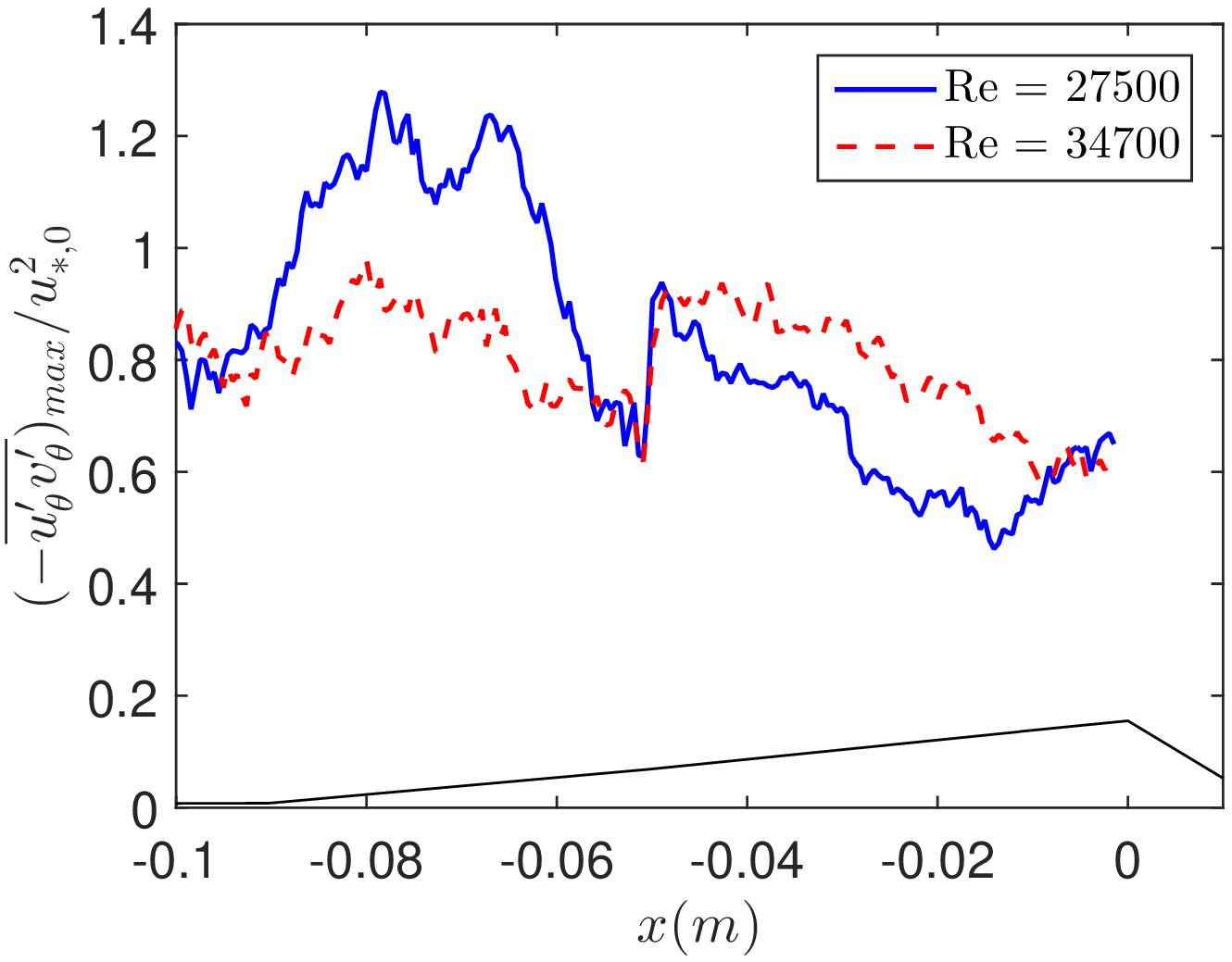}\\
 			(a)
 		\end{tabular}
 	\end{minipage}
 	\hfill
 	\begin{minipage}{0.5\textwidth}
 		\begin{tabular}{c}
 			\includegraphics[width=\textwidth,clip]{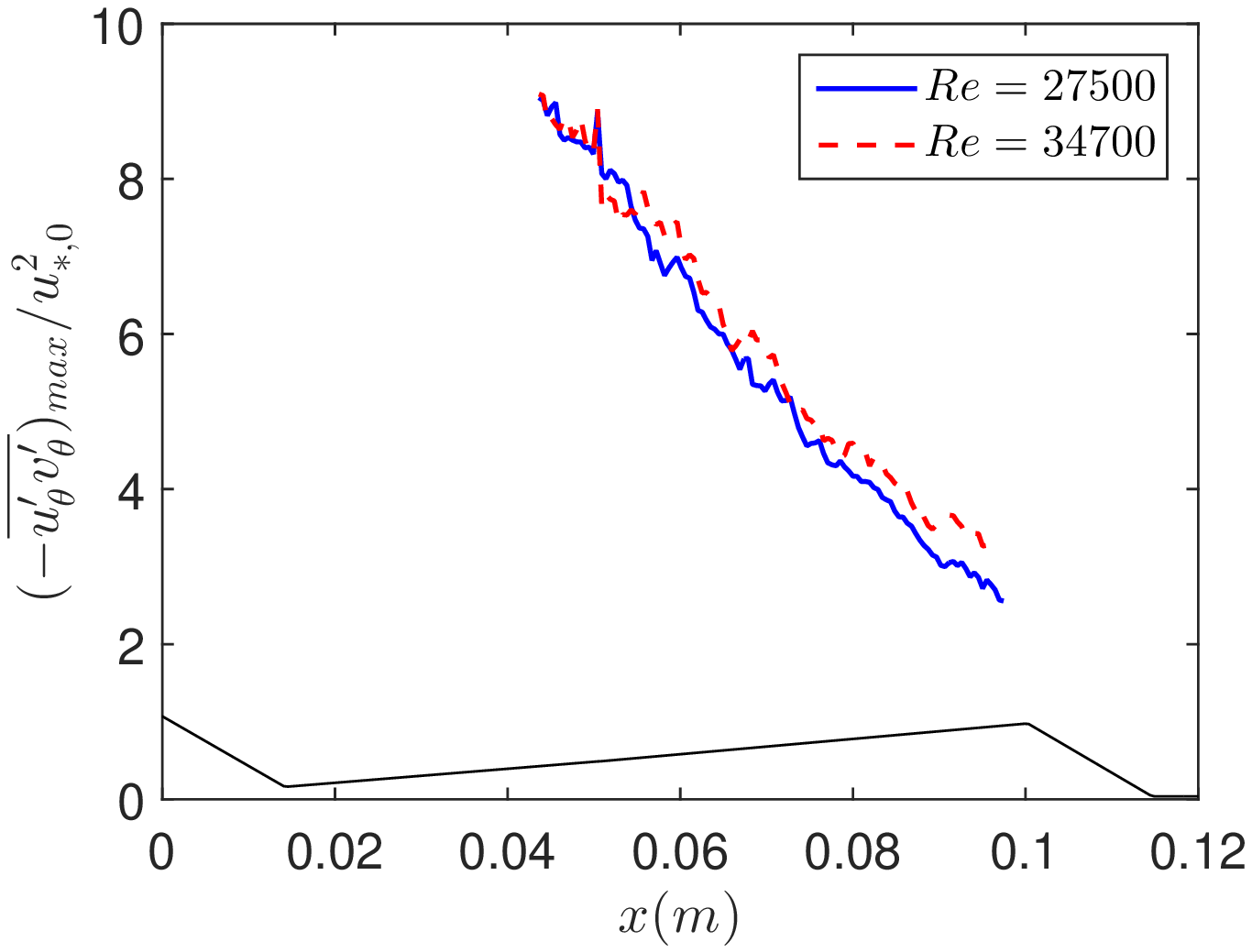}\\
 			(b)
 		\end{tabular}
 	\end{minipage}
 	\caption{Maxima of the normalized Reynolds stress $(-\overline{u'_{\theta} v'_{\theta}})_{max}/({u}_{*,0}^{2})$ as a function of longitudinal position $x$. The blue continuous and red dashed lines correspond to $Re = 2.75\times 10^{4}$ and $Re = 3.47\times 10^{4}$, respectively. The black continuous lines show the ripple height. (a) upstream and (b) downstream ripple.}
 	\label{remax_pair}
 \end{figure}

Figs. \ref{remax}, \ref{remax_pair}a and  \ref{remax_pair}b show the longitudinal evolution of the maxima of the $xy$ component of the Reynolds stress $(-\overline{u'_{\theta} v'_{\theta}})_{max}$ normalized by ${u}_{*,0}^{2}$ for the bottom wall region over the single, upstream and downstream ripples, respectively. The blue continuous and red dashed lines correspond to $Re = 2.75\cdot 10^{4}$ and $Re = 3.47\cdot 10^{4}$, respectively, and Fig. \ref{remax_pair}b presents only the values downstream of the reattachment point. In Fig. \ref{remax_pair}, the black continuous lines show the ripple height. Despite the relatively high noise in the data, some observations can be made.  The profiles for the single and upstream ripples are similar: at $x \approx -0.08$ m (i.e., the beginning of the ripple, or its leading edge), $(-\overline{u'_{\theta} v'_{\theta}})_{max}$ increases by approximately $20\%$, and $(-\overline{u'_{\theta} v'_{\theta}})_{max}$ is greater at some point upstream of the ripple crest. In addition $(-\overline{u'_{\theta} v'_{\theta}})_{max}$ and $u_{*,0}^{2}$ have the same order of magnitude. We note that, concerning the flow over the single ripple, $(-\overline{u'_{\theta} v'_{\theta}})_{max}$ has a peak at $x\, \approx$ -0.03m for $Re = 3.47\cdot10^{4}$. We do not have an explanation for the presence of a peak at this position for this flow condition. The increase of $(-\overline{u'_{\theta} v'_{\theta}})_{max}$ at the leading edge of the ripple can be explained by the presence of a concave curvature in the streamlines of the mean flow. As proposed by Wiggs et al. \cite{Wiggs}, concave curvatures increase turbulent stresses close to walls because turbulent structures from higher-velocity regions, far from the wall, are displaced to near wall regions. On the contrary, near the crest there are convex streamlines, which decrease turbulent stresses close to the crest surface. Finally, because $(-\overline{u'_{\theta} v'_{\theta}})_{max}$ and $u_{*,0}^{2}$ have the same order of magnitude over the upstream face of the single and upstream ripples, equilibrium inner regions can exist upstream of the crest, except at the leading edge and the crest regions; therefore, a shear velocity characteristic of the shear stress at the surface can be computed from the mean profiles.

For the downstream ripple, the longitudinal evolution is completely different. Considering only the values downstream of the reattachment point, the $(-\overline{u'_{\theta} v'_{\theta}})_{max}$ decreases monotonically toward the ripple crest: at the reattachment point $(-\overline{u'_{\theta} v'_{\theta}})_{max}$ is higher than $u_{*,0}^{2}$ by one order of magnitude, and at the crest they are both of the same order of magnitude. Again, this different behavior is due to the flow separation, caused by the upstream ripple, so that local equilibrium is absent in the inner region and a characteristic shear velocity does not exist. We note that the values of $-\overline{u'_{\theta} v'_{\theta}}$ are very high (between approximately 8 and 2 times $u_{*,0}$) close to the surface ($y^+_d$ $<$ 100), and the total surface stress is approximately given by $(-\overline{u'_{\theta} v'_{\theta}})_{max}$ because the mean velocity gradients are not high enough to affect the order of magnitude the total stress.

\begin{figure}[ht]
 	\begin{minipage}{0.5\textwidth}
 		\begin{tabular}{c}
 			\includegraphics[width=\textwidth,clip]{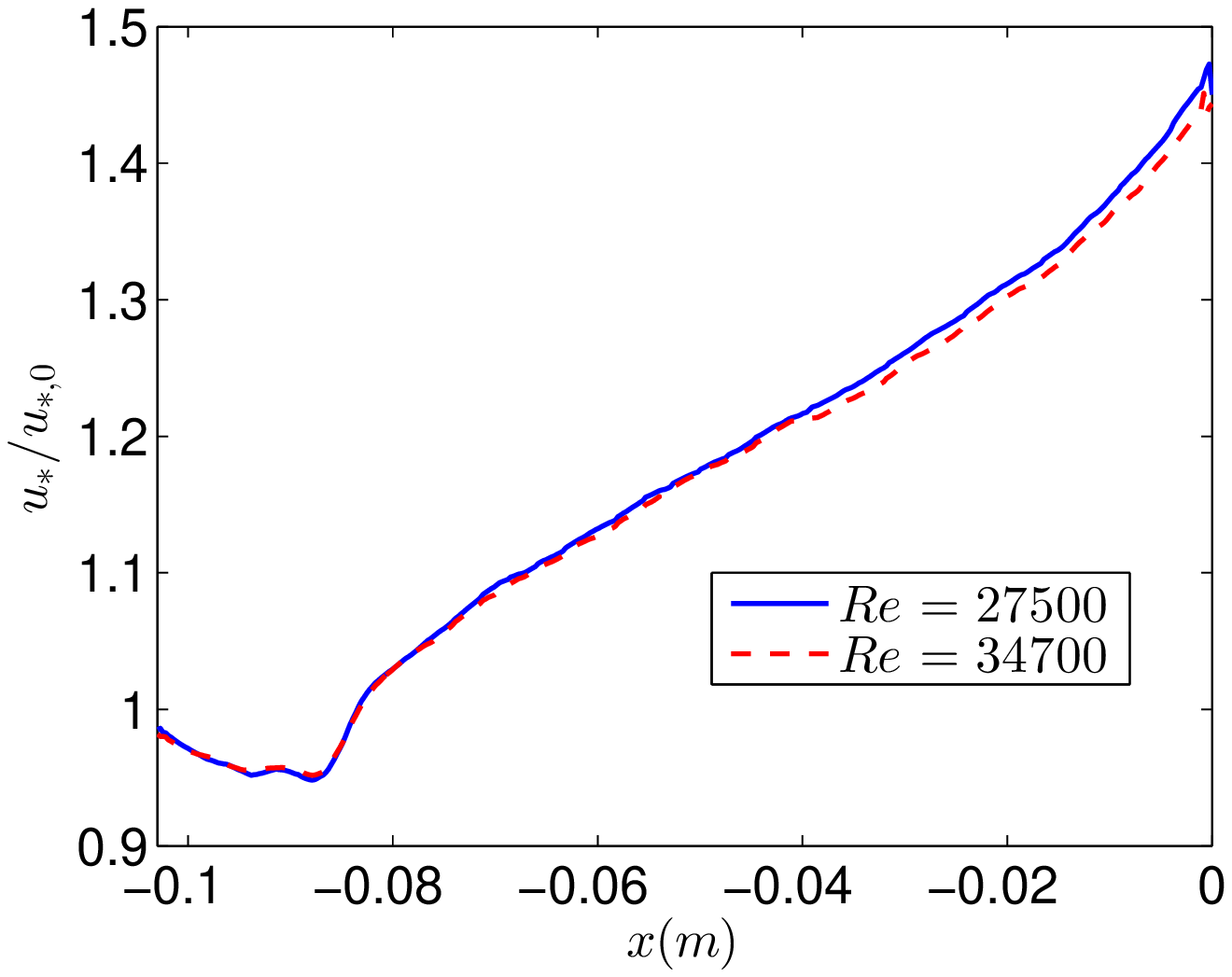}\\
 			(a)
 		\end{tabular}
 	\end{minipage}
 	\hfill
 	\begin{minipage}{0.5\textwidth}
 		\begin{tabular}{c}
 			\includegraphics[width=\textwidth,clip]{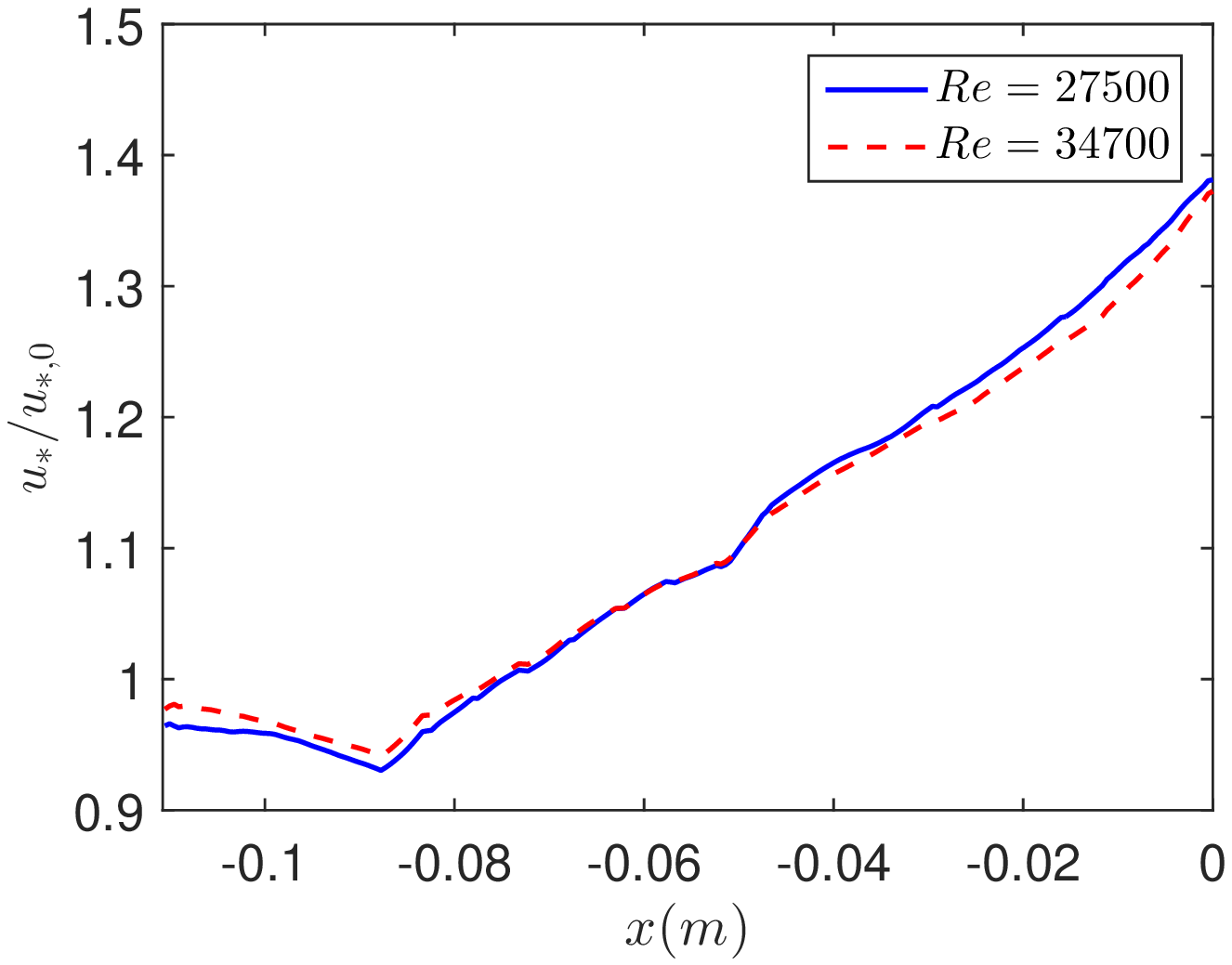}\\
 			(b)
 		\end{tabular}
 	\end{minipage}
 	\caption{Local shear velocity $u_*$ normalized by the unperturbed shear velocity as a function of longitudinal position $x$. The continuous and dashed lines correspond to $Re = 2.75\cdot 10^{4}$ and $Re = 3.47\cdot 10^{4}$, respectively. (a) single ripple and (b) upstream ripple.}
 	\label{local_shear_veloc}
 \end{figure}

Based on the local-equilibrium assumptions for the single and upstream ripples, we computed the local shear velocity for the perturbed flow $u_*$ by fitting each mean velocity profile along the ripple with a logarithmic curve. Figs. \ref{local_shear_veloc}a and \ref{local_shear_veloc}b show $u_*/u_{*,0}$, for the single and upstream ripples, respectively, as a function of the longitudinal position $x$. The continuous line corresponds to $Re = 2.75\cdot 10^{4}$, and the dashed line to $Re = 3.47\cdot 10^{4}$. To eliminate the small fluctuations caused by the Cartesian grid used by the PIV processing software, we averaged the data in Fig. \ref{local_shear_veloc} over the closest nine points using a sliding-window process. We observe that the shear velocity decreases at the beginning of the ripple and then increases to a value approximately 1.4 times $u_{*,0}$. The surface shear stress at the crest of the ripple is approximately twice the value at the channel wall upstream of the ripple. These general behavior and stress values are consistent with measurements of turbulent boundary layers over dunes in deserts, where local-equilibrium conditions in inner regions are evident \cite{Andreotti_1,Sauermann_2,Sauermann_3}.

We note that the maximum shear velocity is not shifted with respect to the ripple crest, in agreement with measurements over aeolian dunes reported in many studies, which also failed to find an upstream shift. In the case of a granular bed, an upstream shift of the surface stress is necessary to explain the growth of sand ripples \cite{Engelund_Fredsoe}. In addition, the presence of lower shear velocities at the leading edge of the ripple would imply local deposition of grains, which would seem to be incompatible with the growth of triangular ripples. However, in the specific case of the leading edge, the relaxation effects related to grain inertia cause erosion in this region \cite{Andreotti_1,Charru_3}. The shear velocity behavior at the leading edge and crest is different from the behavior of turbulent stresses, which increase at the leading edge and exhibit an upstream shift at the crest region. One explanation for this difference is the presence of a concave curvature in the streamlines around the leading edge and a convex curvature around the ripple crest. As proposed by Wiggs et al. \cite{Wiggs}, concave curvatures cause turbulent structures in higher-velocity regions to be displaced to lower-velocity regions, increasing $-\overline{u'_{\theta} v'_{\theta}}$ with respect to $u_*^2$ in the region close to the leading edge, while convex curvatures cause turbulent structures in lower-velocity regions to be displaced to higher-velocity regions, decreasing $-\overline{u'_{\theta} v'_{\theta}}$ with respect to $u_*^2$ in the region close to the crest.

\begin{figure}[ht]
 	\begin{minipage}{0.5\textwidth}
 		\begin{tabular}{c}
 			\includegraphics[width=\textwidth,clip]{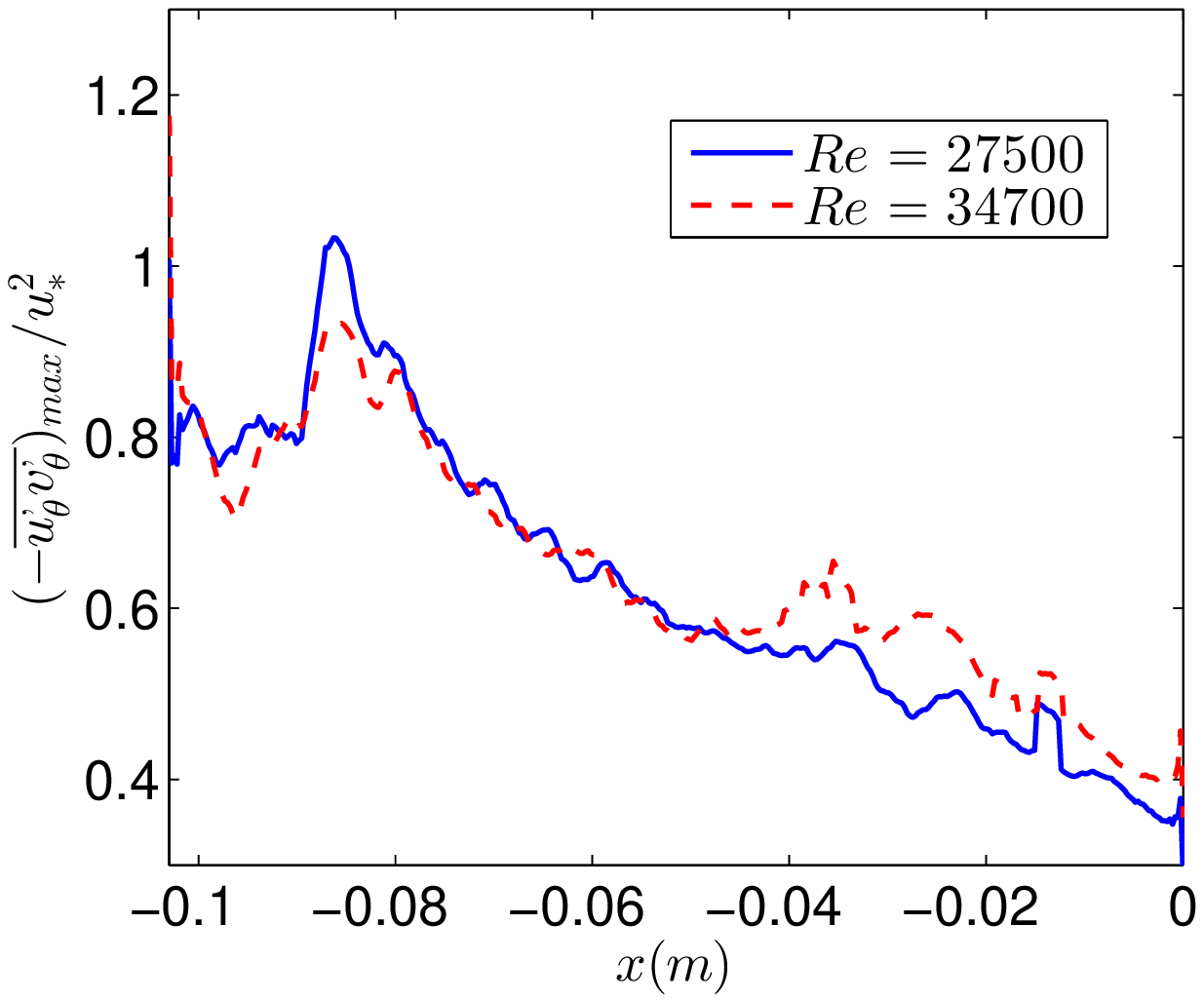}\\
 			(a)
 		\end{tabular}
 	\end{minipage}
 	\hfill
 	\begin{minipage}{0.5\textwidth}
 		\begin{tabular}{c}
 			\includegraphics[width=\textwidth,clip]{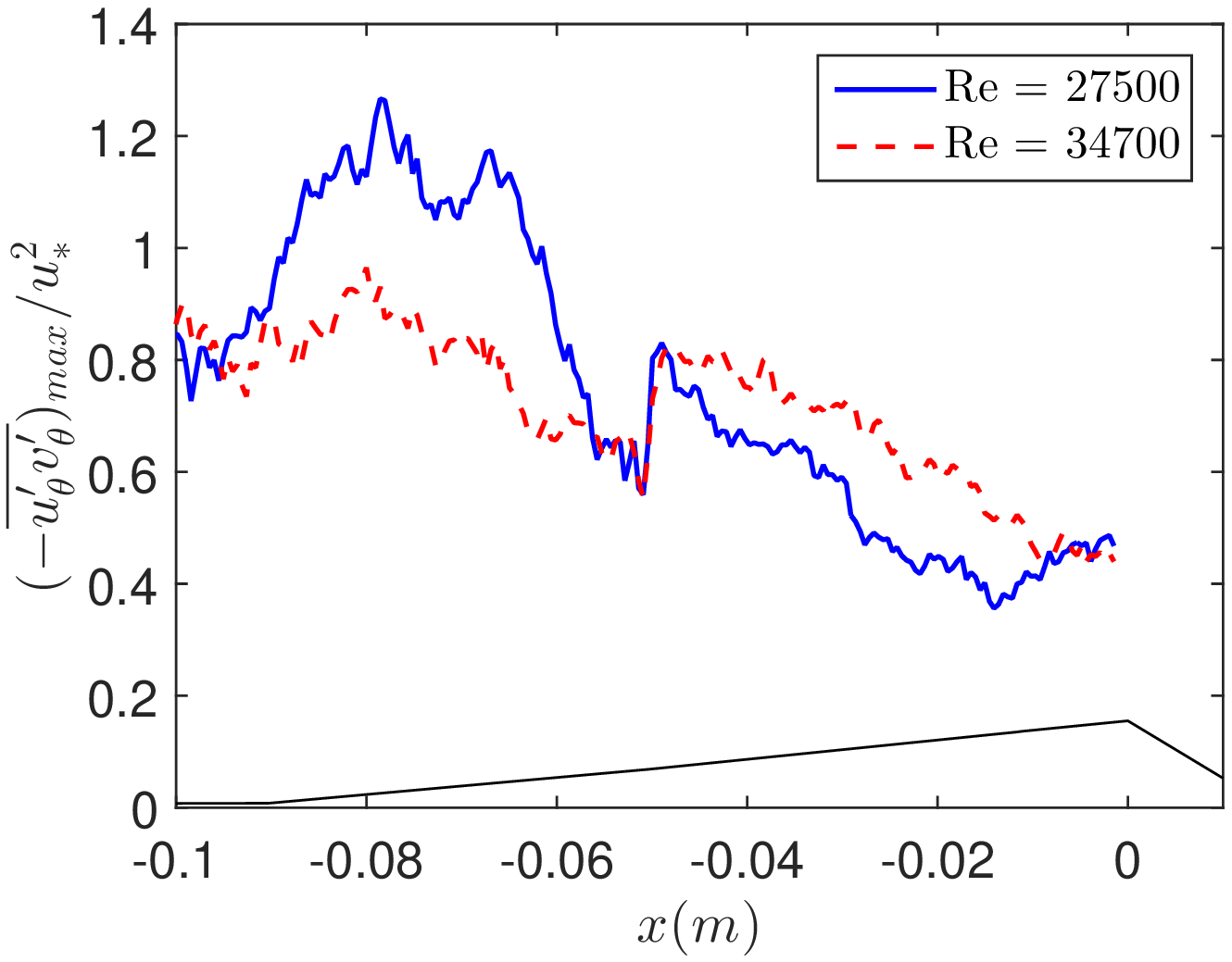}\\
 			(b)
 		\end{tabular}
 	\end{minipage}
 	\caption{Maxima of normalized Reynolds stress $(-\overline{u'_{\theta} v'_{\theta}})_{max}/({u}_{*}^{2})$ as a function of longitudinal position $x$. The continuous and dashed lines correspond to $Re = 2.75\cdot 10^{4}$ and $Re = 3.47\cdot 10^{4}$, respectively. (a) single ripple and (b) upstream ripple.}
 	\label{uv_normal_local}
 \end{figure}

To investigate the flow equilibrium over the ripple, we compared the longitudinal evolution of the maxima of the velocity fluctuations with the longitudinal evolution of the shear velocity from upstream of the leading edge of the ripple to the ripple crest ($x=0$ m), for the single and upstream ripples.  Figs. \ref{uv_normal_local}a and \ref{uv_normal_local}b show the maxima of the normalized Reynolds stresses $(-\overline{u'_{\theta} v'_{\theta}})_{max}/({u}_{*}^{2})$ as a function of the longitudinal position $x$ for the single and upstream ripples, respectively. The continuous and dashed lines correspond to $Re = 2.75\cdot 10^{4}$ and $Re = 3.47\cdot 10^{4}$, respectively. $(-\overline{u'_{\theta} v'_{\theta}})_{max}$ and ${u}_{*}^{2}$ have roughly the same order of magnitude, and $(-\overline{u'_{\theta} v'_{\theta}})_{max}$ has a peak at the leading edge ($x \approx-80 mm$). The $(-\overline{u'_{\theta} v'_{\theta}})_{max}$ then decays gently from the leading edge toward the crest, decreasing more rapidly in the crest region and reaching a value of approximately $40 \%$ of ${u}_{*}^{2}$ at the crest. This is in accordance with the explanation for the fluctuations and mean flow interactions based on the concave/convex curvatures of streamlines proposed in \cite{Wiggs}.

\begin{figure}[ht]
 	\begin{minipage}{0.5\textwidth}
 		\begin{tabular}{c}
 			\includegraphics[width=\textwidth,clip]{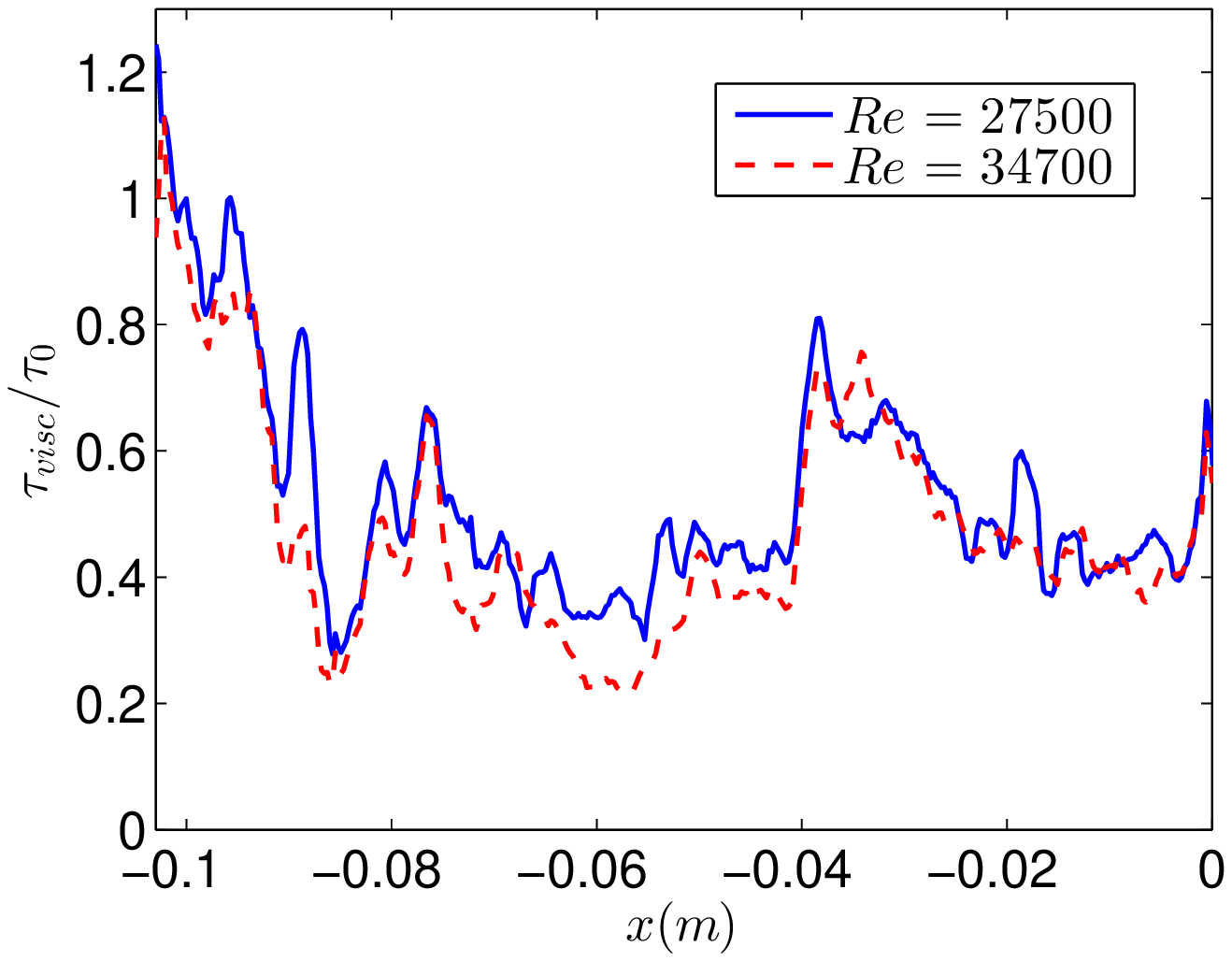}\\
 			(a)
 		\end{tabular}
 	\end{minipage}
 	\hfill
 	\begin{minipage}{0.5\textwidth}
 		\begin{tabular}{c}
 			\includegraphics[width=\textwidth,clip]{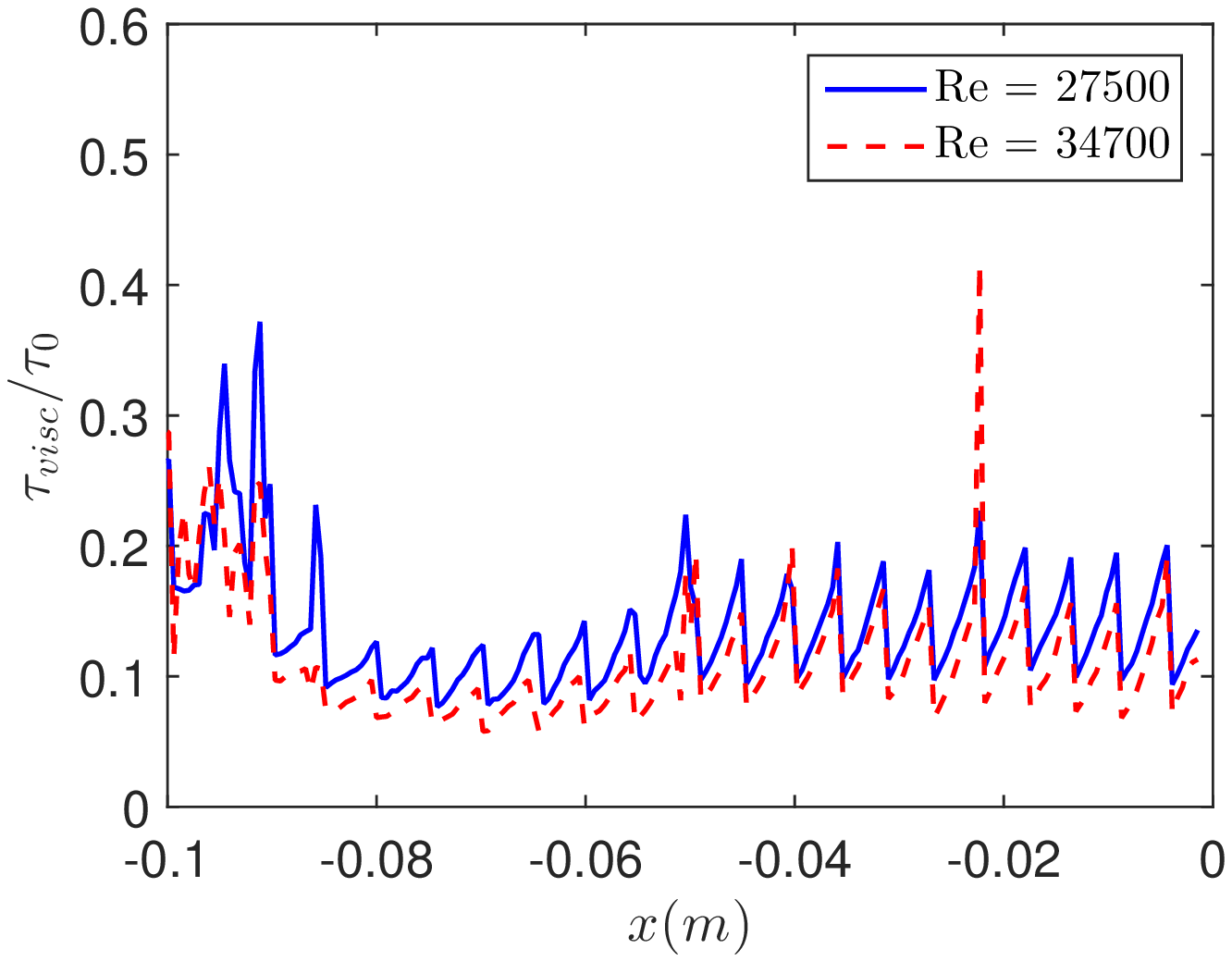}\\
 			(b)
 		\end{tabular}
 	\end{minipage}
 	\caption{Normalized surface viscous stress $\tau_{visc}/\tau_{0}$ as a function of the longitudinal position $x$. The continuous and dashed lines correspond to $Re = 2.75\cdot10^{4}$ and $Re = 3.5\cdot10^{4}$, respectively. (a) single ripple and (b) upstream ripple.}
 	\label{tp}
 \end{figure}

Figs. \ref{tp}a and \ref{tp}b show the evolution along the ripple of the surface viscous stress $\tau_{visc}$ normalized by the unperturbed shear stress $\tau_{0} = \rho u_{*,0}^2$, which was calculated using the Blasius correlation (justified by the results presented in Tab. \ref{tab.1}), for the single and upstream ripples, respectively. The continuous and dashed lines correspond to $Re = 2.75\cdot10^{4}$ and $Re = 3.47\cdot10^{4}$, respectively. Because of the presence of relatively high noise in the numerical computation of gradients, it was not possible to determine the longitudinal position where the maximum surface stress occurs. The high noise is due to the large scatter in the lower part of the boundary layer, caused by undesired reflections from the walls, and to the spatial resolution of the Cartesian grid used for the PIV cross-correlations. The large scatter affects the computation of velocity gradients, and then the viscous stress, at the ripple surface. However, taking into account the relatively high noise, Fig. \ref{tp}a shows that the viscous stress at the upstream surface and $\tau_{0} = \rho u_{*,0}^2$ have roughly the same order of magnitude over a great part of the single ripple. For the upstream ripple, Fig. \ref{tp}b shows lower values of the surface viscous stress because undesired reflections from the walls and a coarser Cartesian grid did not allow us to compute viscous stresses as close to the surface as in the single ripple case.

Information about the local behavior of the flow can be gained from the plot of turbulence production. From the 2D instantaneous fields acquired with the PIV system, the $xy$ component of the turbulence production aligned with the ripple surface is given by:

\begin{equation}
P = (-\overline{u'_{\theta} v'_{\theta}}) \frac{\partial u_{\theta}}{\partial y_{d,\theta}}  
\label{eq4}
\end{equation}

\begin{figure}[h!]
\begin{center}
	\begin{tabular}{c}
	\includegraphics[width=0.55\columnwidth]{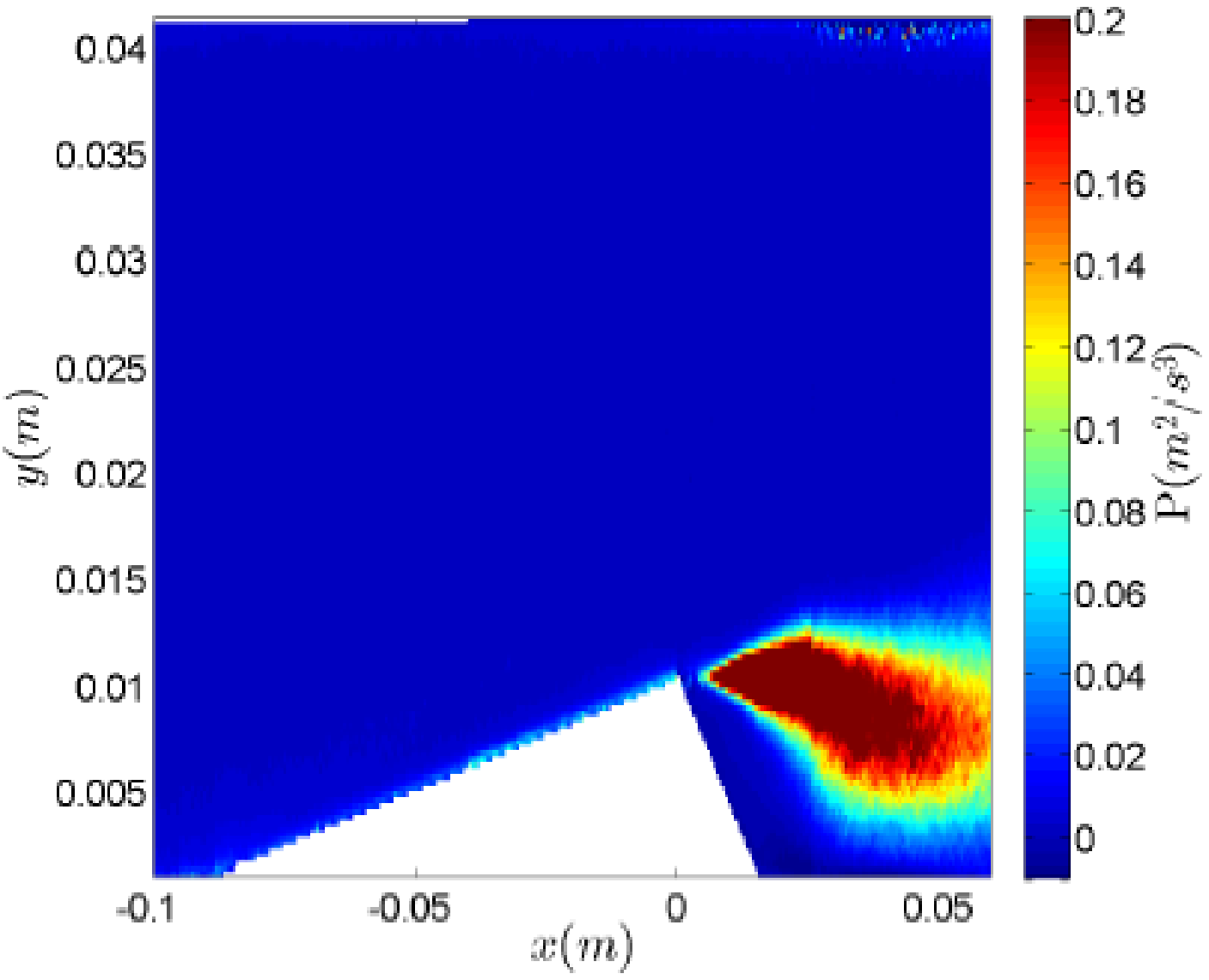}\\
	(a)\\
	\includegraphics[width=0.99\columnwidth]{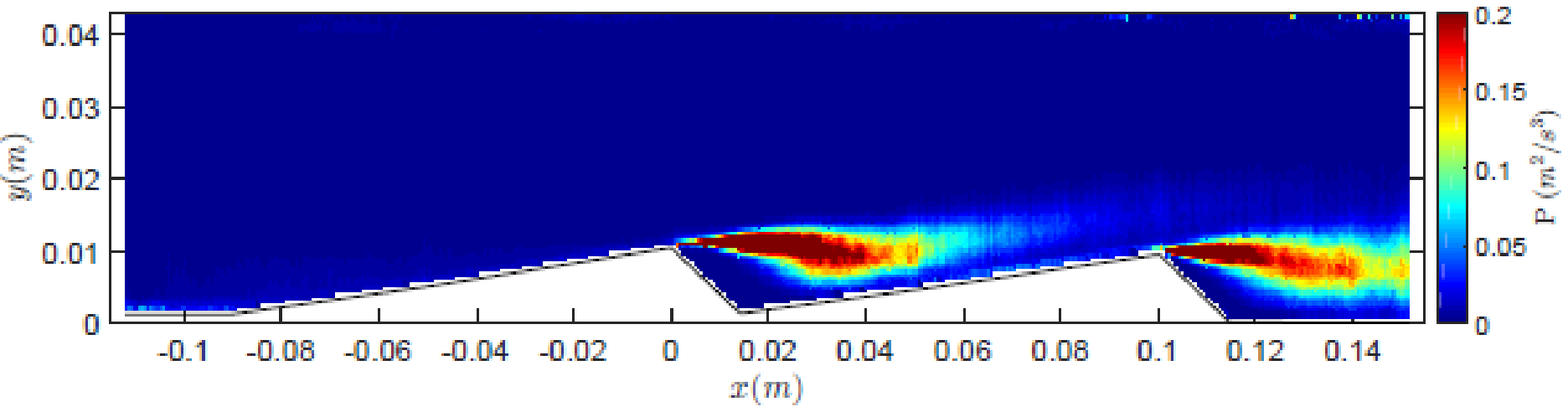}\\
	(b)
	\end{tabular} 
\end{center}
    \caption{Plot of the $xy$ component of turbulence production over the ripple in $x$ and $y$ coordinates for $Re = 2.75\cdot10^{4}$. (a) The single ripple and (b) the pair of ripples.}
    \label{product_ripple}
\end{figure}

Figs. \ref{product_ripple}a and \ref{product_ripple}b show the plot of the $xy$ component of turbulence production in the $x$ and $y$ coordinates for the single and pair of ripples, respectively, and $Re = 2.75\cdot10^{4}$. There are regions of high turbulence production not only in the wake (recirculation bubble), but also the region close to the top wall. The turbulence production close to the top wall is smaller than in the wake, and can be seen in Figs. \ref{product_ripple}a and \ref{product_ripple}b as a thin colored band at their top right region. The production at the top wall occurs downstream of the ripple crest for the singe ripple, within $x>0$ m and $y>0.04$ m, and downstream of the ripple crest of the downstream ripple for the pair of ripples, within $x>0.10$ m and $y>0.04$ m. These higher values found at the upper wall are due to unfavorable pressure gradients downstream of crest of the single and downstream ripples. In the case of the pair of ripples, because there is a stable recirculation bubble between the crests, the unfavorable pressure gradient occurs downstream of the crest of the downstream ripple. The unfavorable pressure gradients generate intense vortices near the crest region on the bottom and top walls, increasing local production of turbulence, as shown in \cite{Laval_1,Laval_2}.

In the region near the crests, the intense turbulence production redistributes fluctuations, changing the relationship between local mean velocities and fluctuations and probably destroying the local-equilibrium conditions. Concerning the single and upstream ripples, this, together with the concave/convex streamlines, explains the absence of an upstream shift for the mean components of the flow, although the fluctuating components exhibit this shift.

\subsection{Discussion: ripples stability}

For the single and the upstream ripples, we note that $(-\overline{u'_{\theta} v'_{\theta}})_{max}$ increases at the leading edge and then decreases at the crest, while $u_*$ decreases slightly at the leading edge and increases toward the crest. This was explained in terms of the curvature of the streamlines \cite{Wiggs}, and implies that turbulent stresses are effectively stronger than expected at the leading edge and weaker than expected at the crest. Therefore, the perturbed boundary layer would appear to be under local-equilibrium conditions along the upstream face of the ripple apart from the leading edge and the crest. Considering the evolution of the total shear stress along the ripple, we observe the existence of an upstream shift between the maximum shear stress and the ripple crest. This is in accordance with asymptotic theories \cite{Jackson_Hunt,Hunt_1,Belcher_Hunt,Carruthers_Hunt,Weng} and allows us to understand the stability of isolated and upstream sand ripples with slightly rounded edges in closed conduits.

For the downstream ripples, the flow is detached and a recirculation bubble exists from the leading edge to a distance of approximately 30\% of the ripple length. From the reattachment point to the crest, the flow evolves differently from the single and upstream cases. In terms of turbulent stress, $(-\overline{u'_{\theta} v'_{\theta}})_{max}$ decreases monotonically from the reattachment point toward the ripple crest, with values at the reattachment point higher than $u_{*,0}^{2}$ by one order of magnitude; therefore, the inner layers are not in local equilibrium. For this reason, there is not a shear velocity characteristic of the shear stress obtained from the mean profiles. However, because of the high values of $-\overline{u'_{\theta} v'_{\theta}}$ close to the surface, the total surface stress is approximately given by $(-\overline{u'_{\theta} v'_{\theta}})_{max}$. In this case, the surface shear stress decreases toward the crest. For a sequence of sand ripples, this means that sand is carried along the upstream surfaces and settles at the crests, explaining why these bedforms exist. The flow over the downstream ripple is different from the single and upstream cases, and is not explained by asymptotic theories \cite{Jackson_Hunt,Hunt_1,Belcher_Hunt,Carruthers_Hunt,Weng}.

Finally, these findings allow us to understand the instabilities of a granular bed and the formation of ripples with slightly rounded edges in closed conduits. While stability analyses usually consider the shear stress evolution given by asymptotic theories \cite{Jackson_Hunt,Hunt_1,Belcher_Hunt,Carruthers_Hunt,Weng}, which are valid for isolated ripples, subaqueous ripples appear in sequence \cite{Franklin_15}; therefore, such theories are valid only for the upstream ripple or for the very early stages of ripples growth. In order to understand the existence and stability of rippled sand beds, we must consider the shear evolution of the downstream ripple.

\section{Conclusions}

This paper has described an experimental study on the perturbation of a closed-conduit flow in turbulent regime by two-dimensional triangular ripples. Two ripple configurations were employed: one single asymmetric triangular ripple, and two consecutive asymmetric triangular ripples, all of them with the same geometry. In the arrangement used here, water flows were imposed in a closed conduit with a rectangular cross section and either one or two ripples fixed to the bottom wall. Just upstream of the single or the pair of ripples, the closed-conduit flow was fully developed and hydraulically smooth, and was considered the unperturbed flow, while the flow over the ripples was considered the perturbed flow. As the blockage ratio over the ripples was around $20\%$, confinement effects were not negligible. To ensure the ripples more accurately represented the ripples found in nature, they had slightly rounded edges. The experiments were performed at moderate Reynolds numbers ($Re$ = ord$(10^4)$). Experimental data for this case are scarce, and the physics is not yet completely understood.

The instantaneous flow fields were measured with PIV, and the results were used to determine the mean velocities and fluctuations for both unperturbed and perturbed flows. From these, the $xy$ components of viscous stresses, Reynolds stresses and turbulence production were computed. The perturbed and unperturbed flows were then compared.
 
The unperturbed flow was found to be fully developed and to follow the law of the wall with a well-defined logarithmic region ($70 < y^{+} < 200$). The $xy$ component of the Reynolds stresses compared well with the shear velocity obtained from the mean profiles, and the friction velocity followed the Blasius correlation.

Upstream of the crests the mean velocity perturbations increase toward the crest and are greater in the region close to the ripple surface ($y_d^+ \lesssim 40$). Far from the ripple surface, the perturbations are weaker and due only to confinement. Downstream of the crests, the flow detaches and a recirculation bubble is generated.

For the single and the upstream ripples, we found that upstream of the crest the 2D component of the Reynolds stresses aligned with the ripple, $-\overline{u'_{\theta} v'_{\theta}}$, is of the same order as the square of the shear velocity of the unperturbed flow, $\rho u_{*,0}^2$, in the $25 < y_d^{+} < 250$ region. In this region, $-\overline{u'_{\theta} v'_{\theta}}$ was of the same order as the viscous stress aligned with the ripple $\tau_{visc} \approx \mu\frac{\partial u_{\theta}}{\partial y_{d,\theta}}$ in the region $y_d^{+} < 25$. These findings indicate that the flow upstream of the crest is in local equilibrium in the inner region. In this case, a logarithmic region in the mean velocity profiles is expected upstream of the ripple crest, and the maxima of the $-\overline{u'_{\theta} v'_{\theta}}$ profiles are good approximations of the local shear stresses on the ripple surface. In addition, the local-equilibrium conditions allow the use of perturbation expressions derived with the mixing-length model.

We computed the local shear velocities $u_*$ along the ripple by assuming local-equilibrium conditions from the flat bottom wall upstream the ripples until the ripple crest. The values of $u_*$ decrease slightly at the leading edge of the ripple and increase significantly toward the ripple crest, reaching a value $40\%$ higher at the crest than the unperturbed value. In addition, we did not observe a shift between the maximum shear velocity and the ripple crest. This behavior is consistent with measurements of turbulent local-equilibrium boundary layers over dunes in deserts.

We showed that $-\overline{u'_{\theta} v'_{\theta}}$ increases and then decreases toward the crest and behaves differently from $u_*$ in the leading-edge and crest regions. The differences in these two regions can be explained by the curvature of the streamlines, which are concave at the leading edge and convex at the crest. This implies that turbulent stresses are effectively stronger than expected at the leading edge and weaker than expected at the crest. In terms of equilibrium, the perturbed boundary layer is under local-equilibrium conditions at the upstream face of the ripple apart from the leading edge and the crest. Analyzing the evolution of the shear stress along the ripple, we observed an upstream shift between the maximum shear stress and the ripple crest. This is in accordance with asymptotic theories \cite{Jackson_Hunt,Hunt_1,Belcher_Hunt,Carruthers_Hunt,Weng} and allows us to understand the stability of isolated and upstream sand ripples with slightly rounded edges in closed conduits.

For the downstream ripple, the flow is detached and a recirculation bubble exists from the leading edge to a distance of approximately 30\% of the ripple length. From the reattachment point to the crest, the flow evolves differently from the single and upstream cases. Individual profiles of $-\overline{u'_{\theta} v'_{\theta}} / u_{*,0}^{2}$ show large peaks at $y^+ \approx 100$, reaching at the reattachment point values more than 8 times that of unperturbed shear stresses; therefore, a region of constant stress within $y_d^+ < 100$ does not exist and the inner layers are not in local equilibrium. For this reason, a shear velocity characteristic of the shear stress obtained from the mean profiles does not exist.

We showed that $(-\overline{u'_{\theta} v'_{\theta}})_{max}$ decreases monotonically from the reattachment point toward the ripple crest, with values at the reattachment point higher than $u_{*,0}^{2}$ by one order of magnitude, and values at the crest of the same order of magnitude as $u_{*,0}^{2}$. Although we were not able to measure accurately the longitudinal evolution of the viscous stress at the ripple surface, the total surface stress is approximately given by $(-\overline{u'_{\theta} v'_{\theta}})_{max}$ because values of $-\overline{u'_{\theta} v'_{\theta}}$ are very high close to the surface ($y^+_d$ $<$ 100). The surface shear stress decreases toward the crest, reaching a peak at the reattachment point since the mean flow is reversed upstream of it. For a sequence of sand ripples, and by considering the inertia of grains, this means that sand is carried along the surfaces upstream of the crests and settles at the crests. This explains the existence of rippled granular beds. The flow over the downstream ripple is different from the single and upstream cases, and is not explained by asymptotic theories \cite{Jackson_Hunt,Hunt_1,Belcher_Hunt,Carruthers_Hunt,Weng}.

Our findings allow us to understand the instabilities of a granular bed and the formation of sequences of ripples with slightly rounded edges in closed conduits. Usually, subaqueous ripples of triangular shape appear in sequence. However, a great part of stability analyses considers the shear stress distribution given by asymptotic theories \cite{Jackson_Hunt,Hunt_1,Belcher_Hunt,Carruthers_Hunt,Weng}, which are valid for isolated ripples, and are applicable only for the upstream ripple or for the very early stages of ripples growth. In order to analyze the stability of rippled sand beds, it is necessary to consider the perturbed flow showed in this paper for the downstream ripple.

\begin{acknowledgements}
Fernando David C\'u\~nez Benalc\'azar is grateful to SENESCYT (Programa de Becas Convocatoria Abierta 2014 Segunda Fase) and to FAPESP (grant no. 2016/18189-0), and Gabriel Victor Gomes de Oliveira is grateful to FAPESP (grant no. 2015/15001-8) and to CAPES for providing financial support. Erick de Moraes Franklin would like to express his gratitude to FAPESP (grant nos. 2012/19562-6 and 2016/13474-9) and CNPq (grant no. 400284/2016-2) for the financial support they provided.
\end{acknowledgements}


\bibliography{references}
\bibliographystyle{spphys}

\end{document}